\documentclass[journal]{IEEEtran}

\usepackage[T1]{fontenc}
\usepackage[utf8]{inputenc}
\usepackage{tipa}
\usepackage{url}
\usepackage{graphicx}
\usepackage{ulem}
\usepackage{subfigure}
\usepackage{amsmath}
\usepackage{mathtools}
\usepackage{afterpage}

\usepackage{cite}

\begin{document}

\title{Steganography in Modern Smartphones and Mitigation Techniques}

\author{Wojciech Mazurczyk and Luca Caviglione
\thanks{W. Mazurczyk is with Warsaw University of Technology (WUT), Institute of Telecommunications, Nowowiejska 15/19, 00-665, Warsaw, Poland, e-mail: wmazurczyk@tele.pw.edu.pl

L. Caviglione is with the Institute of Intelligent Systems for Automation (ISSIA)
of the National Research Council of Italy (CNR), Via de Marini 6, I-16149, Genova,
Italy, e-mail: luca.caviglione@ge.issia.cnr.it.}}

\maketitle

\begin{abstract}
By offering sophisticated services and centralizing a huge volume of personal data, modern smartphones changed the way we socialize, entertain and work. To this aim, they rely upon  complex hardware/software frameworks leading to a number of vulnerabilities, attacks and hazards to profile individuals or gather sensitive information. 
However, the majority of works evaluating the security degree of smartphones neglects steganography, which can be mainly used to: \textit{i}) exfiltrate confidential data via camouflage methods, and 
\textit{ii}) conceal valuable or personal information into innocent looking carriers. 
  
Therefore, this paper surveys the state of the art of steganographic techniques for smartphones, with emphasis on methods developed over the period 2005 to the second quarter of 2014. The different approaches are grouped according to the portion of the device used to hide information, leading to three different covert channels, i.e., local, object and network. Also, it reviews the relevant approaches used to detect and mitigate steganographic attacks or threats. Lastly, it showcases the most popular software applications to embed secret data into carriers, as well as possible future directions. 
\end{abstract}


\IEEEpeerreviewmaketitle

\section{Introduction}
In recent years, the rapid technological advances of software and hardware led to mobile phones offering capabilities previously achievable only for desktop computers or laptops (see reference \cite{smartev} for a discussion of this evolution). The increasing convergence of network services, computing/storage
functionalities, and sophisticated Graphical User Interfaces (GUIs) culminated into new devices called \textit{smartphones}, which are quickly becoming the first choice to access the Internet \cite{traffictrend}, 
and for entertainment. Nevertheless, the Bring Your Own Device (BYOD) paradigm makes them core tools in the daily working routine \cite{byod}. 

This popularity is mainly driven by a multi-functional flavor combining many features, such as a high-resolution camera, different air interfaces (e.g., Bluetooth, 3G and IEEE 802.11), and Global Positioning System (GPS) into a unique tool. To handle the hardware, the typical Operating System (OS) has an architecture very close to the one used on desktops \cite{avslinux}.
Then, the resulting vast user base ignited the development of many applications delivered from online stores, or through specialized sources on the Web. Even if similar hardware/software platforms are at the basis of a more general class of \textit{smart devices} (e.g., tablets), to avoid cannibalization of market shares, many products do not implement the 3G/telephony sub-system. Yet, for the sake of simplicity, we use smartphone as an umbrella term, despite when doubts arise. 

Alas, the unavoidable increasing complexity and the explosion of exchanged data volumes, dramatically multiply vulnerabilities/attacks (see, e.g., references \cite{gensec} and \cite{secsur} for two recent surveys on security issues of mobile devices), or ``social" threats \cite{mioweb2}. For the same reasons, smartphones are becoming an excellent target for profiling users, or to gain sensitive information \cite{steal}. 

In this perspective, \textit{steganography}\footnote{The term derives from the merge of two Greek words, \textit{steganos} ($\sigma \tau \epsilon \gamma \alpha \upsilon$\'{o}$\varsigma$) meaning ``covered", and \textit{graph}\textit{\`i} ($\gamma \rho \alpha \phi \eta$) meaning ``writing". The first evidence of  this practice is rooted back to 440 BC by Herodotus, while the first usage of the term is around 1499. Steganography is sometimes defined as an \textit{art}.} should be explicitly considered to completely understand smartphone security, since it
encompasses various information hiding techniques for embedding a \textit{secret message} into a carefully chosen \textit{carrier}, while cloaking the very \textit{existence} of the communication to any third-party observers. Consequently, steganography is a double-edged sword. On one hand, it can be used by individuals to enforce security and privacy by embedding sensitive data into ``irrelevant" carriers, for instance, images/audio files. On the other hand, it can be exploited by malware to hide the exfiltration of confidential/personal information, or communications towards the Command \& Control (C\&C) in a botnet. 

As today, there are few surveys on information hiding. But, some target a general audience, for instance they describe the history and the development trends of steganography \cite{survey2}, \cite{trends}. On the contrary, other works only focus on specific and narrow areas, e.g., IP telephony \cite{voipsurvey} or stream protocols \cite{sctp}. To our best knowledge, there are no previous comprehensive surveys on information hiding techniques for smartphones, since reference \cite{tesi1} only focuses on colluding applications, and reference \cite{tesi2} is limited to methods based on the physical interaction with the devices, e.g., ``smudge" attacks exploiting oily residues left on the touch screen surface, which are definitely outside the scope of this work. Despite many steganographic techniques to create \textit{covert channels} in desktop-class environments (partially surveyed in \cite{survey2}, \cite{voipsurvey}, \cite{survey1} and \cite{bender}) can be ported to smartphones, the rapid creation of new dedicated approaches imposes to update past efforts.
Having a unique and multi-functional tool combining features previously covered by several separate devices needs a broader investigation. In fact, past surveys could be too ``vertical" in their organization, since they aimed at covering  
specific protocols or services. Instead, smartphones require a wider perspective allowing to grasp all the areas in which information hiding can take place, thus 
demanding for a more ``horizontal" scheme. 
%



Therefore, this paper aims at filling this gap, and its \textit{main contributions} are: \textit{i}) the development of an ``horizontal" taxonomy to classify steganographic techniques for smartphones; \textit{ii}) the systematic review of the literature on information hiding in the period 2005 to the second quarter of 2014, with emphasis on methods used on mobile devices, or that can be successfully borrowed from desktops; \textit{iii}) the evaluation of known or perspective detection approaches, and a quick review of the most popular tools; \textit{iv}) the discussion of the lessons learned, the potential research opportunities, and directions to advance the field of smartphone steganography.

The remainder of the paper is organized as follows: Section \ref{mot} introduces motivations, Section \ref{tsts} outlines the survey architecture, and  Section \ref{sfe} provides the proper background. Then, Section \ref{occ} deals with steganographic methods based on the alteration of digital objects, Section \ref{lcc} to establish a covert channel between two  processes sharing the same physical device, and Section \ref{ncc} to build a flow of information between remote parties. Section \ref{co} reviews the most effective countermeasures, and lastly, Section \ref{cfw} concludes the paper and envisages future research directions. 

\section{Motivations}
\label{mot}

Globally, it is estimated that there are about $5$ billions of mobile phones worldwide, of which $\sim$$1.08$ billion are smartphones \cite{gulf}. Moreover,  according to International Data Corporation (IDC), the worldwide mobile phone market is expected to grow $7.3 \% $ year over year in 2013  \cite{idc}. It must be emphasized that a relevant pulse to this increase is caused by worldwide smartphone shipments that are expected to surpass $1$ billion units for the first time in a single year. For example, in the US alone, in the second quarter of 2012, almost $55\%$ of mobile phone owners had a smartphone, as reported in a survey by Nielsen \cite{nielsen}. 
A key reason of this huge success is the advancement of cellular connectivity, allowing users to interact with high-volume or delay-sensitive services while moving, e.g., through the Universal Mobile Telecommunications System (UMTS), or the Long Term Evolution (LTE) \cite{diffusa}. Proper support of fragmented traffic \cite{wlangame}, jointly with the availability of energy-efficient Graphics Processing Units (GPUs), make them also an excellent platforms for online gaming \cite{smargame}. 

Thus, it is not surprising that smartphones are currently a natural target for steganography, since information hiding techniques were always influenced by how people communicate, also causing steganographic carriers to evolve through the ages \cite{trends}, \cite{stegev}. Accordingly, the research community is even more interested in creating new data hiding paradigms on smartphones, especially for the following reasons: 
\begin{itemize}
 \item \textit{popularity reduces suspicions}: the more a certain carrier is utilized, the better is its masking capacity, as hidden communications can pass unnoticed amongst the bulk of exchanged data;
 \item \textit{``opportunity makes the thief"}: the availability of several applications, services and protocols (possibly combined in a sophisticated manner) dramatically increases the chances for hiding information within the TCP/IP stack, or  smartphone's services, e.g. Multimedia Messaging Service (MMS) or Short Message Service (SMS). Eventually, this makes the smartphone a multi-dimensional carrier. 
\end{itemize}

To demonstrate the importance and timeliness of the topic, we report \textit{four} recent stories on the exploitation of steganography as the enabler to leak information, or to conduct large-scale attacks: 
\begin{itemize}
\item 2008 -- steganographic methods have proven to be useful for data \textit{exfiltration}. It was reported that someone at the U.S. Department of Justice smuggled sensitive financial data out of the Agency by embedding information in several image files \cite{sally};

\item 2010 -- steganography demonstrated that can pass unnoticed for \textit{long periods}. It was discovered that the Russian spy ring of the so-called ``illegals"  used digital image steganography to leak classified information from USA to Moscow \cite{russians}; 

\item 2011 -- steganography witnessed to \textit{scale}. The Duqu worm moved stolen data towards many botnet's C\&C servers (i.e., nodes used for coordination and information gathering purposes) via apparently innocent pictures, thus traversing the Internet without raising any suspicion \cite{duqu};

\item 2014 --  steganography showcased its effectiveness for \textit{signaling}
purposes.  When installed on user's machine, the Trojan.Zbot downloads a 
jpeg file into the infected system. Within the image there is a list of banks and 
financial institutions, which network traffic has to be closely inspected
\cite{zbot}.
\end{itemize}

With such premises, jointly with the rich set of features concentrated within 
an unique device, smartphones enable to use a wide variety of steganographic methods developed in other environments, as well as novel and dedicated techniques. Besides, boundaries among different services and applications are progressively vanishing, thus increasing possibilities of merging different mechanisms into more sophisticated ones. For such reasons, grouping into a unique work the portion of the literature dealing with data hiding in other contexts, but of exceptional interest or straightforwardly applicable to smartphones, along with relevant studies, allow to better understand current scenarios and trends. At the same time, this effort should avoid to underestimate some threats. 

\section{Organization of the Survey}
\label{tsts}

\begin{centering}
\begin{figure}[htbf]
  \center
  \includegraphics[width=8.5cm]{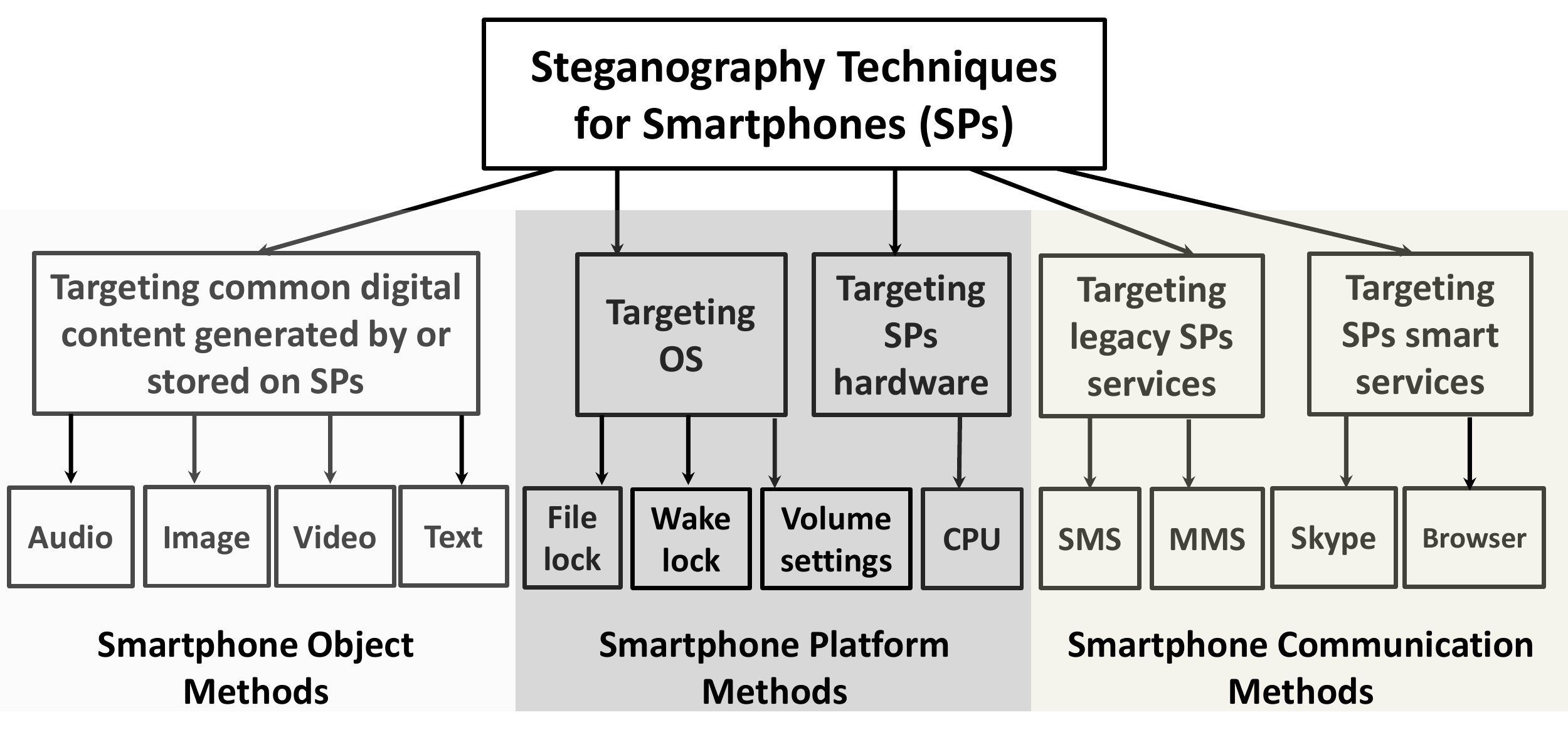}
  \caption{Taxonomy used in this survey to organize the main information hiding techniques targeting smartphone's components and possible examples.}
  \label{taxo}
\end{figure}
\end{centering}

To structure the survey in an efficient way, we group papers (i.e., techniques) according to what component of the device is enabling the steganographic process. To this aim,
we developed the taxonomy depicted in Figure \ref{taxo}, which is composed of three different classes. Each one results into a method and a related hidden channel with a well-defined scope. 
In more details: 

\begin{itemize}
\item \textit{Smartphone Objects Methods} (Section \ref{occ}). The key idea is to use digital objects produced and handled by modern devices as carriers for the hidden data. Methods belonging to this class take advantage of the ``multimedia" vocation of modern smartphones, and can be further partitioned according to the  \textit{type of objects} used, i.e., images, audio, video, and text. 
This section discusses how steganography can be used to embed secrets in objects, which can be remotely sent to create an \textit{object covert channel}, or stored within the device, to increase the safety level of secrets. 
\end{itemize}

\begin{itemize}
\item \textit{Smartphone Platform Methods} (Section \ref{lcc}). As said, modern devices have hardware and software equipments comparable to desktops. Hence, the access to and control of resources and sensitive data require an adequate degree of security, which is usually enforced by dedicated layers within the OS. In this perspective, platform methods are primarily used to bypass the security framework deployed within the smartphone, essentially to enable colluding applications to communicate. In a similar extent, also hardware components can be used to leak information to the surrounding environment: in this case, they will be discussed separately, as to avoid doubts. The resulting fine-grained partition is as follows: 
\begin{itemize}
\item \textit{Targeting OS}: the channel is created by using lower layers of the smartphone's OS, 
such as shared notifications, file-system locks, and manipulable interprocess communication functionalities; 

\item \textit{Targeting Hardware}: the channel lives within the physical behavior of the smartphone, for instance by altering the idle state of the Central Processing Unit (CPU), or its working frequency. Besides, in this class we also present methods using specific peripherals of the smartphone, e.g., the Near Field Communication (NFC) module, the built-in camera or the microphone. 
\end{itemize}
To recap, this section principally reviews methods used to covertly exchange data across different processes within the
same device, thus resulting into a \textit{local covert channel}.
\end{itemize}
\begin{itemize} 
\item \textit{Smartphone Communication Methods} (Section \ref{ncc}). Other than the legacy telephonic communication, smartphones supports a multitude of packet-switched transmission services over the air. Therefore, the resulting channels 
can be subdivided into two main categories: 
\begin{itemize}
\item \textit{Targeting Legacy Services}: the exchange is made by solely using ``telephonic paths", i.e., the data belonging to voice conversations, the SMS or MMS. Many of
the techniques studied in the related literature are adaptations of some objects method to mitigate 
the minimal amount of data embeddable within such vectors. Though, the availability of 
several works offering novel contributions for the case of SMS/MMS make worth mentioning this approach standalone;  

\item \textit{Targeting Smart Services}: the channel is created by virtue of ``Internet-grade" functionalities 
offered by the device, such as a complete TCP/IP protocol stack, as well as the support to full-fledged networked applications (also considering the mobile counterpart of the desktop-class
version). Since investigating steganography for desktop applications is outside the scope of this work, we will only consider here those exploitable in smartphones.
\end{itemize}

Then, in this section we will primarily concentrate on how to take advantage of the communication features of a smartphone  to create \textit{network covert channels}, which are of paramount importance to exfiltrate data outside the device.
\end{itemize}

Nevertheless, the taxonomy will be also used to present techniques to detect, eliminate or limit the aforementioned steganographic channels 
(Section \ref{co}). Similarly, papers and methods will be grouped according to which kind of covert channel they mitigate or prevent (i.e., object, local, and network). 

Despite the proposed organization has been developed to be simple but efficient, some papers mix different techniques simultaneously, thus belonging to multiple classes at the same time. 
When possible, different ideas will be discussed independently in the proper section. Otherwise, the paper will be addressed where it contributes more. 

To avoid overlaps, we will always classify works according to the carrier embedding the secret, rather than how the steganogram is delivered (e.g., a method to hide data into a picture
sent via MMS will be discussed into Section \ref{occ} -- Smartphone Object Methods). Conversely, if the approach deeply exploits features of MMS, it will be discussed into communication methods. 

Although desktop steganography is outside the scope of this survey,
to have the proper foresight, papers discussing techniques, which should/could
appear on smartphones in the near future will be concisely addressed.  
In this case, to avoid confusion, techniques are reviewed separately. \\

To increase the accessibility to the surveyed papers, and to have a quick reference, 
we provide four overview tables: 

\begin{itemize}
\item Table \ref{tabpro} portraits an overall snapshot of the steganographic mechanisms targeting smartphones discussed 
in this survey, as well as a short description of the idea at their basis. Roughly, 
the number of methods are uniformly distributed among the different covert channels, emphasizing the
equal importance of the different carriers. Yet, solutions using SMS/MMS exploit many trivial and limited approaches, thus they should progressively vanish in the near future (at least in this incarnation); 

\item Table \ref{tabpot} groups methods that could/should be ported over smartphones, but 
as today have been only investigated in desktops. As shown, methods relying upon IEEE 802.11 are the most 
promising one, especially by considering the ubiquitous availability of wireless connectivity also in cost-effective devices. Also, the pervasive diffusion of Voice over IP (VoIP)-based services could play a role in the future. For both classes, the major issue preventing their effective utilization relies in software constraints; 

\item Table \ref{tabband} offers a rendition of methods according to their capacity (as defined in Section \ref{dag}). This table only contains works where authors provided a performance evaluation (regardless real or simulated). As it can be noticed, only a small fraction of papers reports quantitative results, or theoretical estimations. Nevertheless, the achieved bandwidth is usually modest, with the exception of techniques using voice codecs; 

\item Table \ref{tabbandpot} is similar to Table \ref{tabband} but focusing on techniques not yet tested in smartphones. Despite the efforts needed to be ported, their adoption over mobile appliances must be further evaluated, since it could reflect in reduced performances imposed by hardware/software limitations, or excessive battery drains. 
\end{itemize}

Lastly, to complete the overview of the major findings of this work: 
\begin{itemize}
\item Table \ref{defin} (Section \ref{sfe}) summarizes all the definitions and properties concerning steganography used through the rest of this survey;

\item Table \ref{tabtools} (Section \ref{sta}) presents a snapshot (updated at February 2014) of the steganography tools available for Android and iOS implementing a modest portion of the surveyed mechanisms. 
\end{itemize}
%

\begin{table*}[t] 
\caption{Steganography methods investigated in the survey classified according to the taxonomy of Section \ref{tsts}.}
\begin{center}
\begin{tabular}{|c|c|c|c|}
\hline
\textbf{Target} & \textbf{CC Type} &  \textbf{Key idea for embedding data} & \textbf{References(s)}\\
\hline
\hline
Image & Object & Data hiding within the LSB of images &  \cite{lsb1}, \cite{lsb2}, \cite{lsb3}\\
\hline
Image & Object & Matrix encoding to reduce modifications & \cite{f5}, \cite{nsf5}, \cite{hugo}\\
\hline
Image & Object & Hidden data within Exif metadata & \cite{exifsteg}\\
\hline
QR Code & Object & DCT/3DES data embedded into a QR Code & \cite{qrcode1}, \cite{qrcode2}\\
\hline
QR Code & Object & LSB/VQ images embedded into a QR Code & \cite{qrcode3},  \cite{qrcode4}\\
\hline 
Audio & Object & Phase coding and echo hiding within audio files  & \cite{bender}\\
\hline
Audio & Object & Spread Spectrum coding within audio files  &\cite{austegs}\\
\hline
Audio & Object & Alteration of bits produced by the Huffman encoder in mp3 & \cite{steglimit}\\
\hline 
Video & Object & LSB alteration applied to video frames & \cite{lsb2} \\
\hline
Video & Object & Embedding/Manipulation of tags in FLV files & \cite{flv}\\
\hline
Text & Object & Semantic alterations of texts sent via mails or chats & \cite{stemail}, \cite{stechat}\\
\hline
Text & Object & Exploitation of metadata within content rich formats, e.g., PDF or Word Documents & \cite{stegword}, \cite{stegpdf}\\
\hline
\hline
SMS & Network/Legacy & Ad-hoc permutations of $3 \times 3$ blocks of pixels in $140$-byte long b/w pictures & \cite{sms11},  \cite{sms12}\\
\hline
SMS & Network/Legacy & Text manipulation, e.g., abbreviations, spaces patterns, or word displacing  & \cite{sms1},  \cite{sms2} \\
\hline
SMS & Network/Legacy & Data posing textual games (i.e., Sudoku) & \cite{sms13} \\
\hline
SMS & Network/Legacy & Patterns based on system/proportional font selection & \cite{sms3} \\
\hline
MMS & Network/Legacy & Spacing in markup languages & \cite{mms1}\\
\hline
MMS & Network/Legacy & Data hiding within the LSB of an image (optimized for MMS) & \cite{mms2} \\
\hline
MMS & Network/Legacy & Alter the quantization step of the Discrete Cosine Transform (DCT) & \cite{mms3} \\
\hline
MMS & Network/Legacy & Apply Pseudo-Random Noise (PRN) on a image (a l\`a spread spectrum) & \cite{mms4} \\
\hline
MMS & Network/Legacy & Embedding of data encrypted with an elliptic curve  &  \cite{mms5} \\
\hline
Voice & Network/Legacy & Manipulation of the Adaptive Multi-Rate Wideband (AMR-WB) codebook bits
used in 3G & \cite{3Gsteg}\\
\hline 
Voice & Network/Legacy & Manipulation of the Adaptive Multi-Rate (AMR) codebook bits 
used in 3GPP standards & \cite{3gppsteg}\\
\hline
Voice & Network/Legacy & Alteration of the last $5$ bits of frames used in the G.723.1 low-rate codec & \cite{gcod}\\
\hline 
Voice & Network/Legacy & Low power tones artificially added within the GSM-encoded voice & \cite{steggsm}\\
\hline
Web/HTTP & Network/Smart & Usage of intentionally crafted URLs & \cite{soundcomber} \\
\hline
Multimedia Stream & Network/Smart & Alteration of the delays between sent video streaming frames &  \cite{gasior} \\
\hline
Web/HTTP/TCP & Network/Smart & Selection of certain content of the banner sent from the remote server & \cite{gasior} \\
\hline
OSN & Network/Smart & Filename alterations and tagging of photos to be posted online & \cite{stegfoto} \\
\hline
OSN/IM & Network/Smart & Attribute manipulation within the XMPP message services & \cite{xmpp} \\
\hline
\hline
OS & Local/OS & Differencing vibration settings  & \cite{soundcomber} \\
\hline
OS & Local/OS & Differencing volume settings  & \cite{soundcomber} \\
\hline
OS & Local/OS & Acquiring and releasing wake-lock permissions & \cite{soundcomber} \\
\hline
OS & Local/OS & Competition for a file lock & \cite{soundcomber} \\
\hline
OS & Local/OS & Modification of single or multiple general settings  &  \cite{collapp} \\
\hline
OS & Local/OS & Automatic broadcasts sent to the interested applications & \cite{collapp} \\ 
&& when modification of particular setting occurs & \\
\hline
OS & Local/OS & Exploitation of type of intents (e.g. flags, action, etc.) in the broadcast messages  & \cite{collapp} \\
\hline
OS & Local/OS & Manipulation of the number of active threads in \textit{/proc} directory  & \cite{collapp} \\
\hline
OS & Local/OS & Using State (open/closed) of a socket & \cite{collapp} \\
\hline
OS & Local/OS & Data coded into number of free blocks on the disk & \cite{collapp} \\
\hline
OS & Local/OS & Exploiting the processor usage statistics available in \textit{/proc/stat} & \cite{collapp} \\
\hline
OS & Local/OS & Utilization of the time elapsed between the sender process changes its priority & \cite{wendzel}\\ && after the screen switches off &  \\
\hline
OS & Local/OS & Usage of time elapsed between the sender process changes its priority  & \cite{wendzel} \\
\hline
Hardware & Local/HW & Hijack of the time elapsed after a screen switch off & \cite{wendzel} \\
\hline
Hardware & Local/HW  & Use of the time elapsed between two consecutive screen activations  & \cite{wendzel} \\
\hline
Hardware & Local/HW & Manipulation of the load on the system and infer bits & \cite{collapp}\\ && by experiencing high or no load in the predetermined time frame &  \\
\hline
Hardware & Local/HW & Variation of  the load on the system and infer bits & \cite{collapp} \\ && by monitoring the trend of the processor frequency & \\
\hline
\end{tabular}
\end{center}
\label{tabpro}
\end{table*}


\begin{table*}[t]
\caption{Steganography methods potentially portable over smartphones and major challenges to be addressed.}
\begin{center}
\begin{tabular}{|c|c|c|c|c|}
\hline
\textbf{Target} & \textbf{CC Type} &  \textbf{Key idea for embedding data} & \textbf{Major Challenges} &\textbf{Ref.(s)}\\
\hline
\hline
Video & Obj. & Alteration of color space of an MPEG stream & Not known if the hardware is able to extract the hidden  & \cite{mpeg}\\
&&& data without impairments  or excessive battery drains & \\
\hline
Video & Obj. & Alteration of motion vectors of an MPEG stream & Not known if the hardware is able to extract the hidden & \cite{video}\\
&&& data without impairments  or excessive battery drains & \\
\hline
Text & Obj. & Alteration of metadata within {\tt .doc} or {\tt .pdf} documents & Compatibility tests to reveal possible limitations &
\cite{stegword} \\
& & & in the implementation on smartphones & \cite{stegpdf}\\
\hline
\hline
NIC & Net./HW & Alteration of the Cyclic Prefix of IEEE 802.11n frames & Hard to deploy for desktops and smartphones & \cite{ofdm}\\
 &&&  since the access to hardware/NIC driver is required &\\
\hline
NIC & Net./HW & Manipulation of fields of the FCF of IEEE 802.11a &  Hard to deploy for desktops and smartphones &\cite{80211steg}\\ &&&  since the access to hardware/NIC driver is required &\\
\hline 
NIC & Net./HW & Exploitation of bits used in padding of IEEE 802.11g & Hard to deploy for desktops and smartphones &\cite{wipad}\\
&&&  since the access to hardware/NIC driver is required &\\
\hline 
NIC & Net./HW & Hiding data in well-suited corrupted frames &  Hard to deploy for desktops and smartphones & \cite{hiccups}\\
&&&  since the access to hardware/NIC driver is required &\\
\hline
\hline
Skype & Net./SS & Utilizing PDUs with silence to carry secret data bits & Some difficulties may arise due to & \cite{skyde}\\
&&& limited capabilities/functionality of mobile OSs & \\
\hline
VoIP & Net./SS & Voice transcoding to free space for secrets & Some difficulties  may arise due to limited & \cite{transteg}\\ &&&  capabilities/functionality of actual mobile OSs & \\
\hline
VoIP & Net./SS & Exploitation of intentionally highly delayed voice PDUs &  Issues  may arise due to limited & \cite{lack1}\\ &&&  capabilities/functionality of actual mobile OSs & \cite{lack2}\\
\hline
\hline
BitTorrent & Net./SS & Manipulates the order of PDUs in the BitTorrent protocol & Some difficulties may arise due to &  \cite{stegtorrent} \\ &&& limited capabilities/functionality of mobile OSs & \\
\hline
\hline 
Quake III & Net./SS & Manipulates some bits of the angles used to & Tests are needed to quantify if the overhead can be  & \cite{q3a}\\
&& locate the player on the game world &  faced by the HW already running the 3D engine &\\
\hline

\end{tabular}
\end{center}
\label{tabpot}
\end{table*}

\begin{table*}[t]
\caption{Steganography bandwidth of surveyed methods (as/if provided in the reference paper).}
\begin{center}
\begin{tabular}{|c|c|c|c|}
\hline
\textbf{CC Type} & \textbf{Manipulated Carrier}  & \textbf{Bandwidth [bps]} &  \textbf{Ref.}\\
\hline
\hline
Local/OS & Vibration Settings & $87$ & \cite{soundcomber} \\
\hline
Local/OS & Volume Settings & $150$ & \cite{soundcomber} \\
\hline
Local/OS & Screen State & $5.29$ & \cite{soundcomber}\\
\hline 
Local/OS & File Locks & $685$ & \cite{soundcomber}\\
\hline
Local/OS & General Settings & $56$ (single),  $260$ (multiple) & \cite{collapp} \\
\hline
Local/OS & Automatic Intents  & $71$ & \cite{collapp}\\
\hline
Local/OS & Automatic Intents  & $3,837$ & \cite{collapp}\\
\hline
Local/OS & Threads Enumeration  & $148$ & \cite{collapp}\\
\hline
Local/OS & UNIX Socket Discovery  & $2,129$ & \cite{collapp}\\
\hline
Local/OS & Free Space on Disk  & $11$ & \cite{collapp}\\
\hline
Local/OS & Reading {\tt /proc/stat}  & $5$ & \cite{collapp}\\
\hline
Local/OS & Task List/Screen State & $0.3$ & \cite{wendzel}\\
\hline
Local/OS & Process Priority & $0.3$ & \cite{wendzel}\\
\hline
Local/OS & Process Priority & $17.6$ & \cite{wendzel}\\
\hline
\hline
Local/HW & CPU & $3.7$ & \cite{collapp}\\
\hline
Local/HW & CPU & $4.9$ & \cite{collapp}\\
\hline
Local/HW & Screen State/Elapsed Time & $0.22$ & \cite{wendzel}\\
\hline
\hline
Network/SS & Live streaming/Altering delays & $8$ & \cite{gasior2}\\
\hline
\hline
Legacy/Voice & AMR-WB codec & $1,400$ & \cite{3Gsteg} \\
\hline
Legacy/Voice & AMR codec & $2,000$ & \cite{3gppsteg}\\
\hline
Legacy/Voice & G.723.1 codec & $133.3$ & \cite{gcod}\\
\hline
\end{tabular}
\end{center}
\label{tabband}
\end{table*}

\begin{table*}[t]
\caption{Steganography bandwidth of methods potentially portable over smartphones (as provided in the reference paper).}
\begin{center}
\begin{tabular}{|c|c|c|c|}
\hline
\textbf{CC Type} & \textbf{Manipulated Carrier}  & \textbf{Bandwidth [bps]} &  \textbf{Ref.}\\
\hline
\hline
WLAN/HW & IEEE 802.11 (cyclic prefix) & $3.25$ M (BSPK), $6.5$ M (QPSK), &  \cite{ofdm}\\
&& $13.0$ M (16-QAM), and $19.5$ M (64-QAM) & \\
\hline
WLAN/HW & IEEE 802.11 (FCF) &
$16.8$ & \cite{80211steg}\\
\hline
WLAN/HW & IEEE 802.11 (padding) &  $1.1$ M for data frames, $0.44$ M for ACKs &
\cite{wipad}\\
\hline
WLAN/HW & IEEE 802.11 &  $216$ k & \cite{hiccups}\\ 
\hline
\hline
Network/SS & Skype packets with silence &  $2,000$ & \cite{skyde}\\ 
\hline
Network/SS & VoIP stream payload &  $32$ k & \cite{transteg}\\ 
\hline
Network/SS & Chosen voice packets' payload &  $600$ & \cite{lack1}, \cite{lack2} \\ 
\hline
\hline
Network/SS & BitTorrent peer-peer data exchange protocol &  $270$ & \cite{stegtorrent}\\ 
\hline
\hline
Network/SS & Quake III Protocol & $8 - 18$ (more will cause lags in the game) &
\cite{q3a}\\
\hline
\end{tabular}
\end{center}
\label{tabbandpot}
\end{table*}

\vspace{0.5cm}
\section{Steganography Fundamentals and Its Evolution}
\label{sfe}
In this section we provide a compact tutorial on steganography and its recent evolution. We recall, that major definitions/properties used throughout the rest of the 
survey are summarized in Table \ref{defin}.

\begin{table*}[t]
\caption{Definitions and Properties of Steganography used in the Survey.}
\begin{center}
\begin{tabular}{|c|c|}
\hline
\textbf{Term} & \textbf{Definition} \\
\hline
\hline
Steganography & The set of information hiding techniques for \\ & embedding a secret message into a well-suited carrier\\
\hline
Steganographic (or Hidden Data) Carrier & The  innocent looking carrier (e.g., a picture) embedding the secret message\\
\hline
Secret Message, Hidden Message & The message to be sent covertly also maintaing third-party observers unaware\\
\hline
Steganogram & The result of secret data embedded within the carrier \\
\hline
Covert Channel & The resulting secret channel (also called the steganographic channel) \\
\hline
Overt Channel & A channel that is not hidden and publicly observable\\
\hline
Steganalysis & The pool of techniques for the detection of a covert communication \\
\hline
\hline
\textbf{Property} & \textbf{Definition}\\
\hline
\hline
Steganographic Bandwidth (or Capacity) & The volume of secret data that can be sent per time unit \\ & for a given steganographic method \\ 
\hline
Undetectability (or Security) & The inability to detect a steganogram within a carrier\\
\hline
Robustness & The amount of alterations a steganogram can withstand without \\ & destroying the embedded secret data\\
\hline
\hline
\textbf{Type of Steganographic Channel} & \textbf{Definition} \\
\hline
\hline
Object Covert Channel & The channel is created by exchanging secrets embedded in
a digital object \\
\hline
Local Covert Channel & The channel has a scope limited within the physical boundaries of the device \\
\hline
Network Covert Channel & The channel exists to exfiltrate data from/to the device\\
\hline
\end{tabular}
\end{center}
\label{defin}
\end{table*}%

\subsection{Definitions and Goals}
\label{dag}
Steganography is the process of embedding secret messages into an innocent looking carrier, defined
as \textit{steganographic} or \textit{hidden data carrier}. Its importance increases with the ubiquitous availability of digital information, since the procedure to inject data is influenced by how communication methods evolve, e.g., from letters written with sympathetic ink, to computer networks. 
When used to conceal information in digital environments, the 
scientific community has been using many denominations, mainly steganography \cite{survey2},  \cite{fridrich}, covert channel \cite{lampson}, \cite{dod}, or information hiding \cite{survey2}. This stems from the introduction of terms 
in different \'epoques, and reflects the vitality of this research field. We point out that, this work adheres on the belief that \textit{``steganographic methods are used to create covert (steganographic) channels"} \cite{commag}. 

Typically, steganography is used to achieve one of the following goals: 
\begin{itemize} 
\item to conceal a valuable information in a carefully chosen carrier to keep a secret safe;  
\item to hide the very existence of the communication if the carrier is transmitted between the parties involved in steganographic data exchange. Eventually, this further permits to keep any third-party observers unaware of its presence.
\end{itemize}

In both cases, choosing the most suitable data carrier is crucial to guarantee an adequate degree of secrecy. The two most important properties for a carrier are: 

\begin{itemize}
 \item \textit{popularity}: the used vector should be not considered as an anomaly itself, potentially unmasking the existence of the hidden communication; 
 \item \textit{unawareness}: modifications needed to embed a secret data should be not ``visible" to a third party unaware of the steganographic procedure.
\end{itemize} 
\indent
\textit{Example:} consider a steganographic method hiding data through the manipulation of a digital image. Its alteration would lead to artificial noise, potentially revealing the steganographic communication. Therefore, distortions (noise) introduced by the steganographic procedure must be not perceptible by human senses, i.e., the visual system.\\		

The inverse of steganography is defined as \textit{steganalysis}, and it concentrates on the detection of a covert  communication. Such a discipline has started to surface fairly recently, mainly ignited by developments in
the performance of hardware and availability of ad-hoc software tools. Its modern nature reflects in applications for the embedding of data considerably outnumbering those dedicated to the detection and extraction of secrets.  
For instance, as February 2014, the largest commercial database of steganographic tools (not limited to smartphones) contains more than $1,200$ applications \cite{sarc}, while Hayati et al. \cite{steganalysis} only mention $111$ tools dedicated to steganalysis (alas, the last update is dated back  to 2007).

\subsection{Model and Scenarios}
\begin{centering}
\begin{figure}[t]
  \center
  \includegraphics[width=8.5cm]{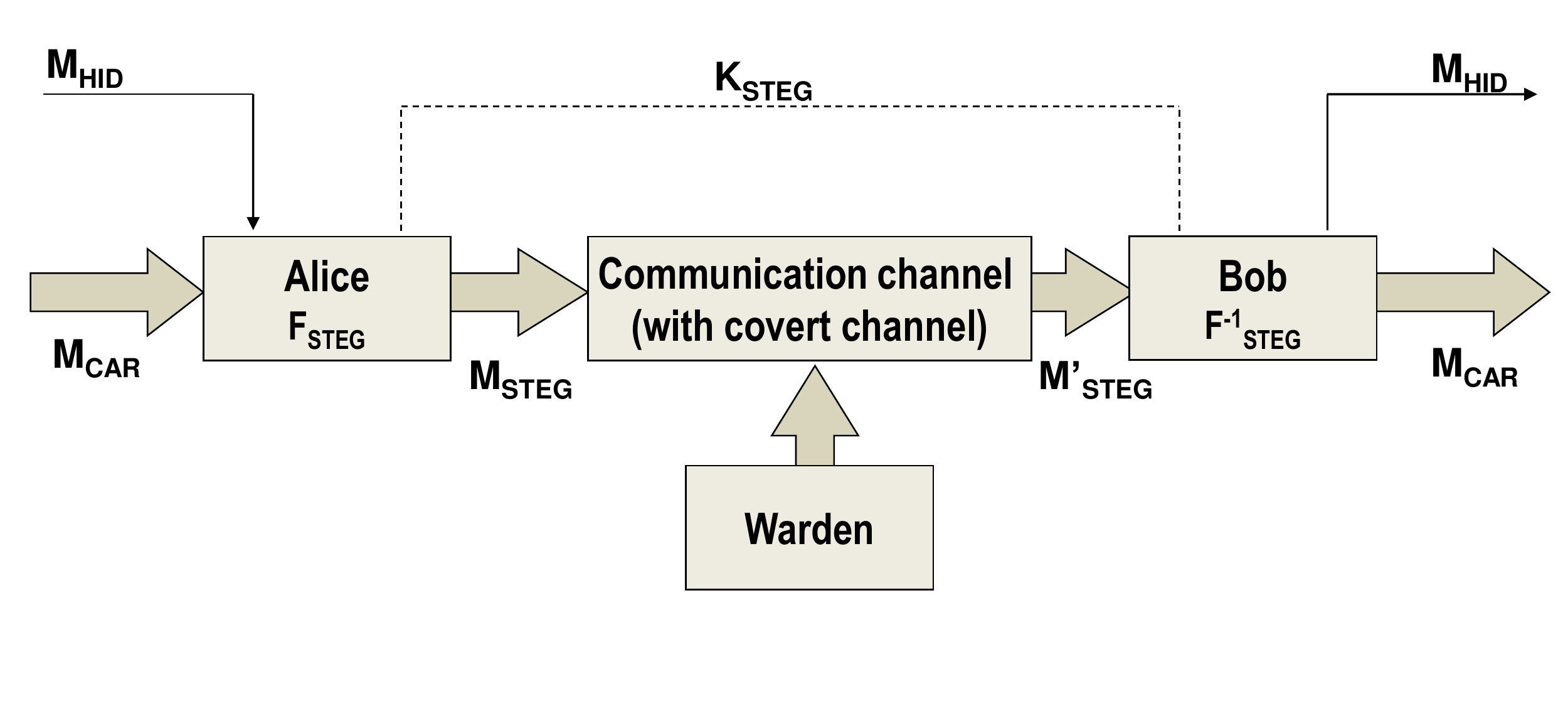}
  \caption{Model for hidden communication based on the Simmons' prisoners problem \cite{simmons}. }
  \label{model}
\end{figure}
\end{centering}

The hidden communication model underlying steganography is based on the famous “prisoners' problem” \cite{simmons}, firstly formulated by Simmons in 1983, and graphically represented in Figure \ref{model}. 

The model considers two prisoners, Alice and Bob, jailed in separate cells, and trying to 
communicate to prepare an escape plan. Their messages are always passed through -- 
and inspected by the Warden: if any conspiracy is identified, he will put them into solitary confinement. 
To succeed in the escape, Alice and Bob must find a way to exchange hidden messages. 

To be more formal, denote as $M_{HID}$ the hidden message, and as $M_{CAR}$ the innocent looking carrier.  To bypass the Warden, both  Alice and Bob must agree on a steganographic method: this is the pre-shared knowledge denoted as $K_{STEG}$. Then, to hide and unhide the secret, they use  $F_{STEG}(\cdot)$ and $F^{-1}_{STEG}(\cdot)$, respectively. To create a covert communication channel, 
%
%
%
the sender alters $M_{CAR}$ by applying $F_{STEG}(\cdot)$ to have a $M_{STEG}$:
\[ 
M_{STEG} = F_{STEG} \bigl ( M_{CAR}, M_{HID}\bigr)  
\] 

Similarly, the receiver will use $F^{-1}_{STEG}(\cdot)$ with  $M_{STEG}$ to have $M_{HID}$, as well as the original $M_{CAR}$:
\[
\bigl ( M_{HID},  M_{CAR} \bigr)  = F^{-1}_{STEG} \bigl( M_{STEG} \bigr)
\]

However, the channel (the Warden in the scenario of Simmons) may alter the hidden message $M_{STEG}$, for instance due to noise, resulting into $M^{'}_{STEG}$. This can be also the outcome of an intentional data morphing/alteration to impede steganography (see, e.g., \cite{tmor1} for a detailed discussion on preventing network  traffic camouflage). As a consequence, the steganographic communication could be voided due to difficulties in computing the $F^{-1}_{STEG} (\cdot)$. 

\begin{centering}
\begin{figure}[t]
  \center
  \includegraphics[width=8.5cm]{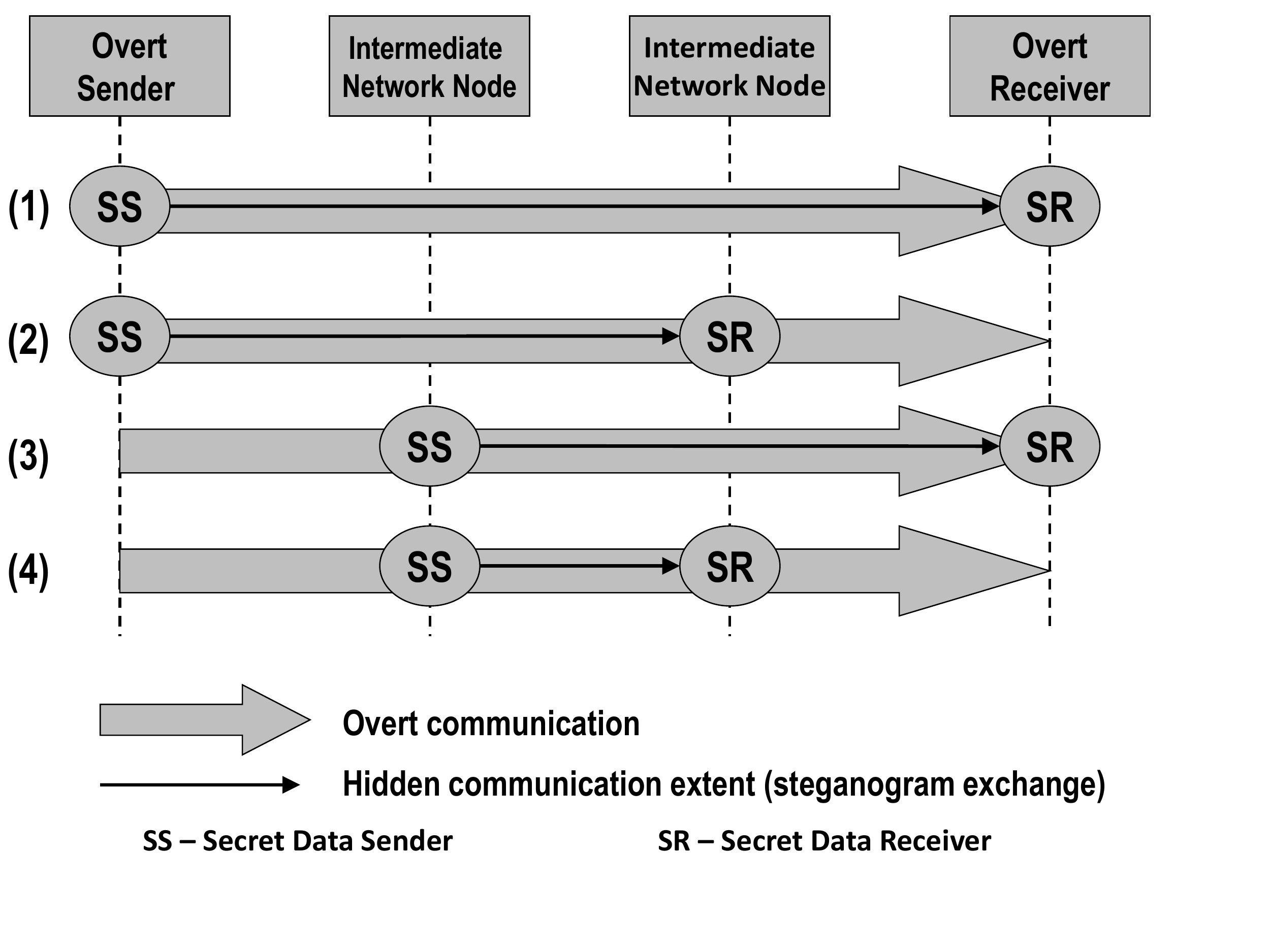}
  \caption{Hidden communication scenarios for steganography generalizing covert/overt channels. }
  \label{scenarios}
\end{figure}
\end{centering}

The development of digital communications enabled steganography to be used in more general forms than the one originally envisaged by Simmons. As shown in Figure \ref{scenarios}, a secret exchange can be implemented in \textit{four} 
different scenarios, labeled as (1) -- (4).  
By denoting with {\tt SS} the sender of secret data, and with {\tt SR} the receiver, two core usage patterns 
are possible:
\begin{itemize}
\item [ ] (1): {\tt SS} and {\tt SR} perform some \textit{overt} data exchange, while simultaneously exchanging secret data. As a consequence, the overt  and the hidden paths are the same; 
\item[ ] (2) -- (4):  only a part of the overt end-to-end path is used for the hidden communication by {\tt SS} and {\tt SR}, which are located within intermediate nodes. Thus, the sender and receiver are, in principle, unaware of the steganographic data exchange.
\end{itemize}  

\subsection{Fundamental Properties}
All steganographic methods exchanging data between two endpoints can be characterized by the following properties (also defined as performance indexes): 
\begin{itemize}
\item \textit{steganographic bandwidth} (or \textit{capacity}): it is the volume of secret data  sent per time unit by using a given method;
\item \textit{undetectability} (or \textit{security} \cite{fridrich}): it is the inability to detect a steganogram within a certain carrier (e.g., by comparing the statistical properties of the captured data with the typical known values for the carrier);
\item \textit{robustness}: it is the amount of alterations a steganogram can withstand without destroying the embedded secret data. \\
\end{itemize}

Unfortunately, steganographic properties heavily depend on the specific methods, and their computation via formal frameworks is still an open research problem \cite{stemod}. Therefore, they are often evaluated via empirical analyses or simulation tools. 

\begin{centering}
\begin{figure}[t]
  \center
  \includegraphics[width=5cm]{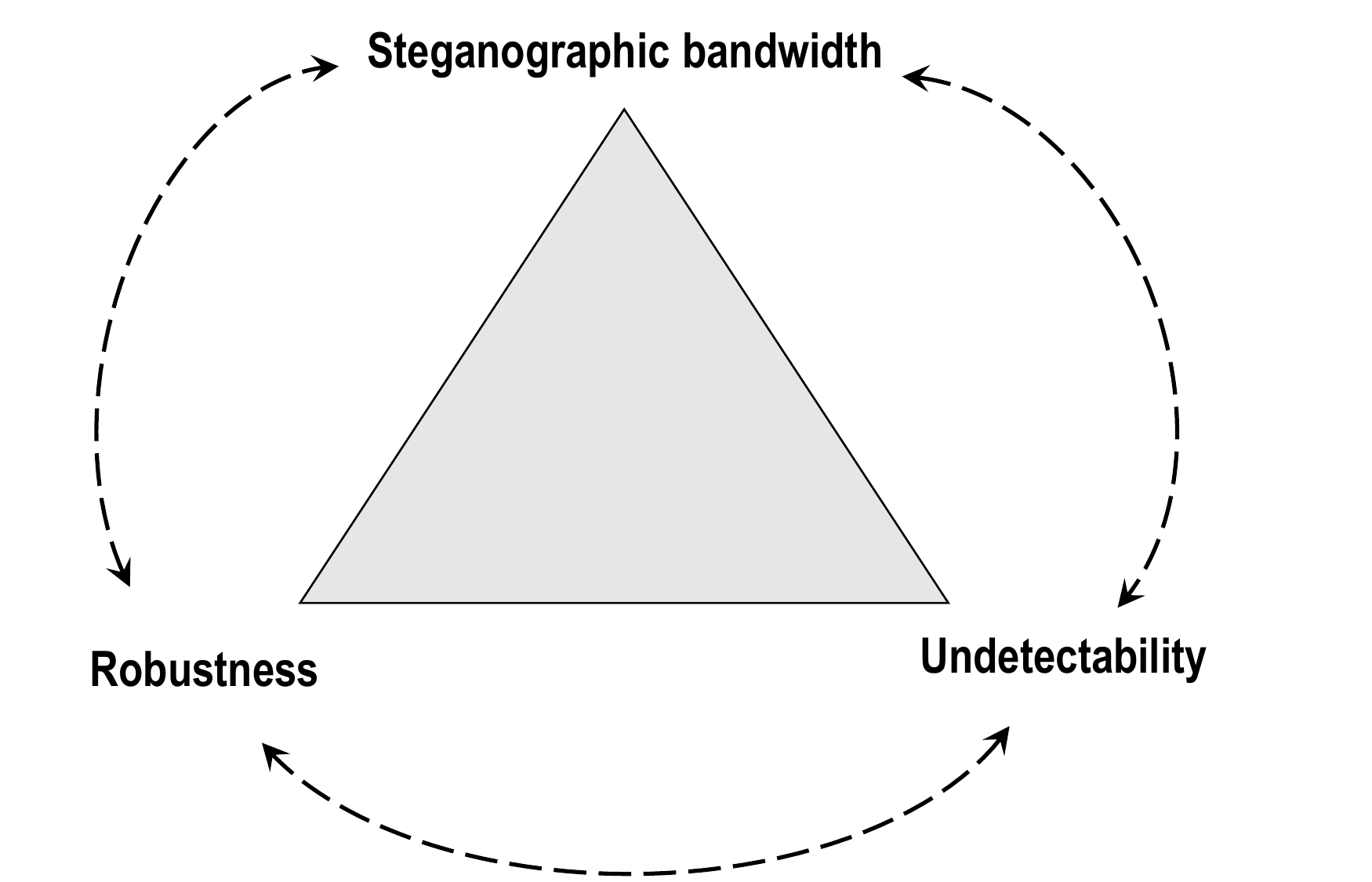}
  \caption{The trade-off relation between the three performance indexes characterizing a covert channel \cite{triangle}. }
  \label{triangle}
\end{figure}
\end{centering}

Ideally, the perfect steganographic method should be both robust and hard to detect, while offering the highest possible bandwidth. But, a trade-off among the performance indexes is always necessary, as thoroughly discussed in the seminal work of Fridrich \cite{triangle}. The relation ruling the interdependence of the metrics is  often called \textit{magic triangle}, and depicted in Figure \ref{triangle}. Put briefly, its rationale is that
is not possible to increase a performance index without lowering the other two. 
For the sake of completeness, we just mention the \textit{magic hexagon}, which describes a more enhanced space \cite{hexagon}, but it is seldom adopted. 

\subsubsection*{Example - Applying the magic triangle to image steganography} consider using some bits of an image as a method to embed secrets. Alterations result into noise, which must be kept under a given threshold in order to be undetected. Yet, the more bits are 
used for the secret (the bandwidth increases), the more would be the noise, thus making the method 
more fragile (detectable) \cite{triex}.  

\subsection{Evolution of Digital Steganography}
\label{eds}
We briefly review the most relevant techniques for the evolution of steganography, spawned by 
the advent of digital information and networking. 
In this perspective, the cutting-edge research on data hiding for
smartphones focuses on:  
\begin{itemize}
\item \textit{digital media};  
\item \textit{linguistic} (or \textit{text});
\item \textit{device resources}; 
\item \textit{network}.  
\end{itemize}

We point out that \textit{linguistic} techniques are not dynamically evolving as others, but the complete support by modern smartphones of textual communication and software for editing documents make them relevant. 

\subsubsection{Digital media steganography} it is the most established part of the information hiding discipline, as it dates back to the 1970's, when researchers developed algorithms to secretly embed a signature in digital pictures (called a \textit{watermark}). During the years, different methods have been proposed, ranging from Least Significant Bit (LSB) modifications, to texture block coding \cite{bender}. Modern smartphones can also manipulate video, thus starting the creation of many 
techniques to embed data within an Moving Picture Experts Group (MPEG) video carrier \cite{mpeg}, e.g., secrets can be stored within video artifacts, or in the metadata describing the compressed media \cite{video}. 
A similar approach is used for audio, since the human auditory system can be eluded through its ``masking property", i.e., noise or stronger tones make weaker ones inaudible \cite{aumask}. Besides,
smartphones are also digital jukebox, thus many works consider audio-specific compressed formats, like MPEG - Layer III (mp3) in a way similar to the video counterpart (see, e.g., reference \cite{mp3steg} for a survey on the topic).

\subsubsection{Linguistic (or text) steganography} it exploits various aspects of the written word, hence the carrier is the syntactic and/or the semantic structure of the text. The most popular techniques are: word-spacing alteration, displacement of punctuation marks, word order manipulation or patterns built on the choice of synonyms \cite{stetext}. Currently, owing to the large volume of daily spam messages, an emerging field investigates their adoption for steganographic purposes \cite{spam}. 

\subsubsection*{Object Covert Channel}
techniques that belong to \textit{1}) and \textit{2}) allow to create an \textit{object covert channel}  that can be further utilized to \textit{i)} conceal the users' secrets with the intent of storing them locally on the device, or \textit{ii)} if a modified file is sent to a party aware of the applied steganographic scheme, to hide the very existence of the communication. 

\subsubsection{Device Resources Steganography} it embraces techniques to form a covert channel within the same physical entity. Device resources steganography includes methods that typically allow inter-processes data exchanges within the single host. The quintessential carrier is a system-wide resource, e.g., the CPU load \cite{cpu}, the level of a buffer, memory zones, or storage resources. Such techniques have been recently becoming en vogue, since they are increasingly applied to achieve steganography in cloud computing environment \cite{cloud}. 

\subsubsection*{Local Covert Channel} methods using this kind of steganography enables to 
create a local hidden path among entities within the same device.

\subsubsection{Network Steganography} it is the youngest branch of information hiding, but it is the most fast developing. Network steganography may utilize one or more protocols simultaneously, for instance by exploiting relationships between different layers of the ISO/OSI stack. One of the earliest mechanism, considered as the very ancestor, is the utilization of different fields (mainly unused or reserved bits) within the header of protocols belonging to the TCP/IP suite as a hidden data carrier \cite{rowland}. 
More advanced methodologies include steganography in real-time services like IP telephony \cite{transteg}, or in peer-to-peer (p2p) services such as, Skype \cite{skyde} or BitTorrent \cite{stegtorrent}. Recently, Online Social Networks (OSNs) like Facebook \cite{stegobot} have been targeted as a possible carriers of hidden data. Nevertheless, specific techniques exploiting emerging network protocols, like the Stream Control Transmission Protocol (SCTP) \cite{sctp} are part of the ongoing research. 

\subsubsection*{Network Covert Channel} network steganography techniques allow to create a channel for communicating in a hidden manner from/to a device, and typically this is done through a communication network.

\subsubsection*{Remark} in digital media steganography the aim was to \textit{fool the human senses} as to make the distortions introduced by applying information hiding technique not perceptible by a person. On the contrary, in local resources and network steganography the key goal is to \textit{fool other processes or machines}. As a consequence, steganographers must properly consider the totally different characteristics of the adversary.

\section{Smartphone Objects Methods}
\label{occ}
As discussed, modern devices can capture and manipulate audio and video, take high-resolution pictures and create text via full-featured word processors. Additionally, the Software as a Service (SaaS) paradigm allows to edit contents even in presence of resource limited devices. As briefly introduced in Section \ref{eds}, smartphones offer a variety of carriers to embed secret data, being a relevant platform for exploiting the so called \textit{digital media steganography}.

Potentially, all the methods developed for desktops could be applied to 
mobile platforms, and many techniques have been also offered to consumers via online stores (a list dedicated to smartphone is compiled in Section \ref{sta}). However, many of the available solutions are plain ports, and the related literature  introduces little novelty, with the only notable exception of steganographic approaches based on Quick Response (QR) codes.

Therefore, the papers presented in this section are limited to objects/techniques that has been successfully tested within a mobile device.

\subsection{Images}The most popular technique used for embed secrets in digital images is still the well-known LSB algorithm \cite{bender}. As the name suggests, it relies on modification of least significant bits of pixels, and it has been proposed in many variations \cite{lsb1}, \cite{lsb2} and \cite{lsb3}, and tools (e.g., MobiStego, Stego Message, Incognito, and Steganographia, to mention some). Yet, LSB steganography is widely considered as a poor choice, since it can be easily nullified by introducing random substitutions in the least significant bits of pixels values. 

One exception is given by the more sophisticated F5 algorithm \cite{f5} (for instance, implemented by the Pixelknot software). F5 has a high steganographic capacity and it is quite hard to detect, since it utilizes matrix encoding to improve the efficiency of the embedding procedure, and to reduce the number of required image modifications. Also, it applies permutative straddling to uniformly scatter the secret data across the carrier image, thus increasing its robustness. Even if many improvements to F5 have been proposed recently, e.g., nsF5 \cite{nsf5} or HUGO \cite{hugo}, they are still not publicly available to users or other researchers. 

Lastly, the increasing diffusion of the Exchangeable image file format (Exif), containing metadata
and information such as tags, and GPS positions, makes also possible to use it as a carrier to store hidden information 
(see, e.g., \cite{exifsteg}). 

\subsection{QR codes}
\begin{centering}
\begin{figure}[h]
  \center
  \includegraphics[width=2cm]{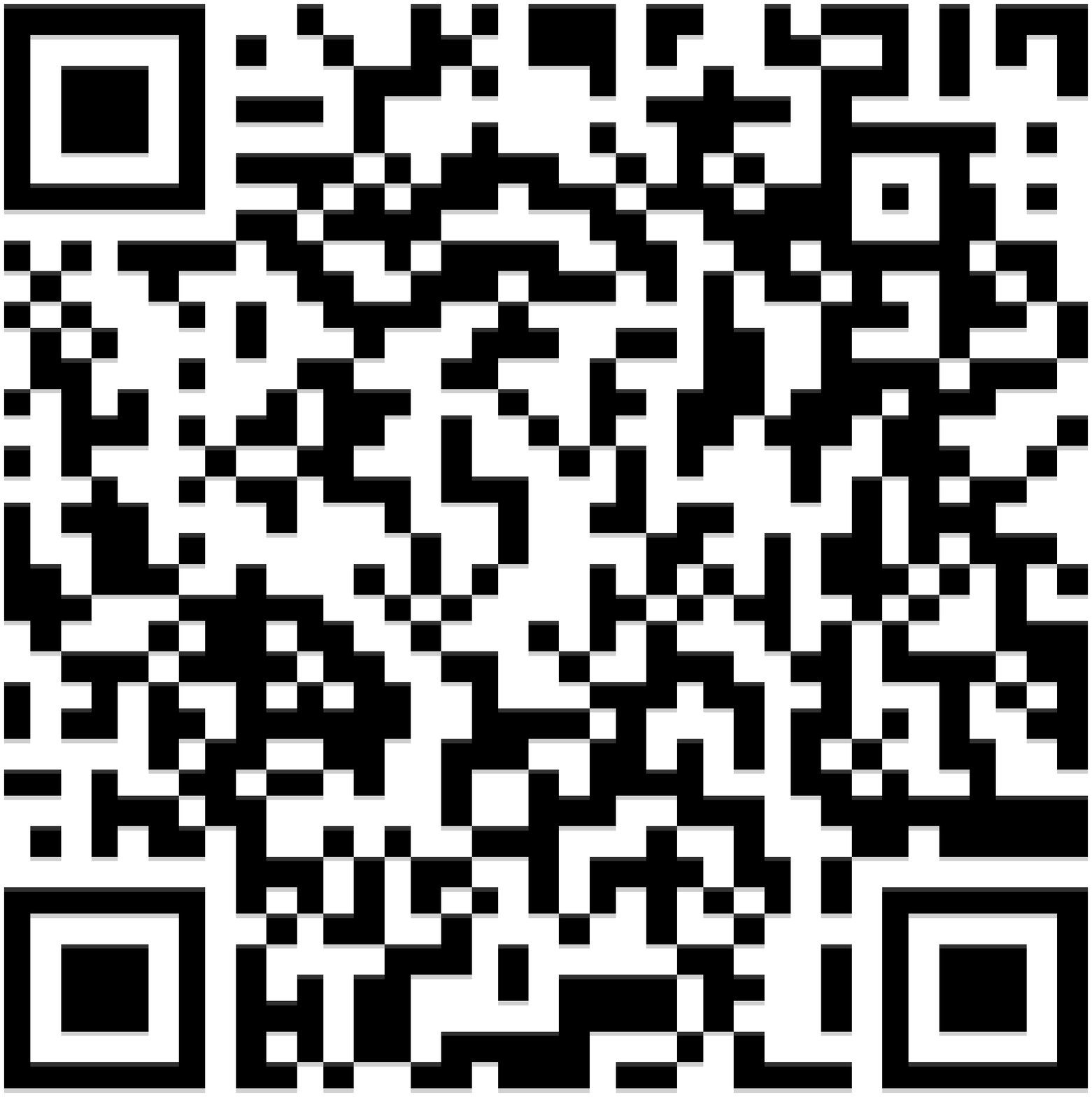}
  \caption{Example of QR code.}
  \label{qrcode}
\end{figure}
\end{centering}

QR codes are barcodes with black and white square dots resembling random noise (an example of QR code is presented in Figure \ref{qrcode}). Despite other contactless technologies like the Radio-Frequency IDentification (RFID), a QR code does
not require dedicated readers, or the need of deploying physical tags. QR codes can be created in a very simple way, for instance by using free software tools. Therefore, they appear in many places, such as posters or flyers, and they can store simple text, contact information, or URLs. To access the information, it is sufficient to capture the code with a camera, which is decoded via a proper application.   

From a steganographic viewpoint, QR codes can be also adopted as hidden data carriers. In this vein, they are used as a ``first step" before applying some digital image steganographic technique. Specifically, Chung et al. \cite{qrcode1} propose to transform secret data into a QR code, and then embed it in a digital image by performing the Discrete Cosine Transform (DCT) which is a commonly used technique to decorrelate the image data. 
Similar approaches are discussed in \cite{qrcode2} and \cite{qrcode3}: the former utilized QR code encryption using 3 DES (Data Encryption Standard) for higher security and then the LSB to embed resulting data into a covert image. The latter combines the Discrete Wavelet Transform (DWT) and the Singular Value Decomposition (SVD) into a DWT-SVD method to produce a stealth QR code within a picture. 

Another application is proposed by Wu et al. \cite{qrcode4}, where authors rely on steganography to 
hide the presence of the QR code for aesthetic purposes. Specifically, they use edge detection and
Vector Quantization (VQ) to make a QR code invisible by humans, as to hide its presence in images or commodities. 

\subsection{Audio}
As hinted, one of the most relevant portion of research to embed secretes into audio contents is based on the elusion of the human auditory system. Besides classical LSB methods (working with the same principle of those used for image steganography, where instead of pixels, audio samples are manipulated), the most popular algorithms use phase coding, echo hiding \cite{bender} and Spread Spectrum (SS) \cite{austegs}. Specifically: 
\begin{itemize}
\item \textit{phase coding}: the phase of an initial audio segment is substituted with a reference phase that represents the secret data. To protect the relative phase between subsequent segments the phase of the segments is adjusted; 

\item \textit{echo hiding}: to hide data, synthetic resonances, crafted in the form of closely-spaced echoes are introduced. The latter,  being very close in the frequency domain, are inaudible by the human ears (this technique is also exploited by the StegDroid tool);

\item \textit{SS}: originally proposed as a steganographic tool in \cite{vbased}, it scatters the secret data across the entire spectrum of the audio carrier. Typically, it offers a lower steganographic capacity compared to LSB, but it is significantly harder to detect. 

\end{itemize}

While the previous techniques exploit the audio in a non-compressed form, the ubiquitous diffusion of the {\tt .mp3} format spawned several dedicated approaches, where secret data is embedded by using the random alterations of the parity bits produced by the Huffman encoder \cite{steglimit}. 

Yet, compression could be also an unwanted effect. For instance the original audio embedding data could be compressed by a third part, thus destroying the created steganogram. In this perspective, Bao and Xiaohu \cite{mp3bis} discuss an mp3-resistant method to protect the secret data by means of a well-suited \textit{a-priori} manipulation of the dynamic range of the uncompressed audio object (similar to the pre-emphasis process used to compensate attenuation in analog telephony).  

Also, by virtue of the wide literature on digital processing to face noise, today is now well understood that error correction coding is a good supplemental carrier for audio steganography, since any redundant information can be adopted to convey secret information, even at the cost of losing some robustness to random errors \cite{bender}. 

\subsection{Video}

The exploitation of video content as a hidden data carrier is a rather new approach if compared with image or audio steganography. Yet, the underlying techniques mostly overlap, since a video can be demuxed into an audio stream and a sequence of images. Besides, smartphones surely enable 
to capture videos, but processing and post-production phases are operations best accomplished on desktops. Therefore, by surveying the literature on video steganography for smartphones, we found that: \textit{i)} there are not many papers nor tools dealing with data hiding within videos produced by smartphones; \textit{ii)} a number of works tend to focus on techniques that previously have been exploited for other digital media. Relevant examples are tools like CryptApp or Steganography Application. Both use the LSB algorithm to hide secret on a per-frame basis. A similar approach is showcased in the work by Stanescu et al. \cite{lsb2}.  

An important idea is to exploit the structure of specific file formats for video contents. In this vein, Mozo et al. \cite{flv} investigate the alteration of Flash Videos ({\tt .flv}), which are becoming
widespread owing to the popularity among users of sites like YouTube. In details, authors propose to  embed  secrets by modifying video tags or by intentionally deleting tags/fields used to populate the metadata structure for describing/organizing the contents.

To summarize, the engineering of novel approaches targeting video steganography for smartphones is quite an unexplored field, hence reflecting in a lot of room for new research. 

\subsection{Text}
Since smartphones can be used for creating text, we  mention here \textit{text steganography}, which can be classified into \textit{three} main branches \cite{textsteg}: 

\begin{itemize}
\item \textit{format-based}: it modifies a text to hide data. Typical techniques are: well-defined misspellings, fonts resizes, and spaces/line skipping patterns. Even if such methods can be hardly detected by a human,
they produce severe statistical alterations easily recognized by machines. For such reasons, they are anachronistic and seldom used, mainly by unskilled users in emails and SMS (see, Section \ref{lst});
 
\item \textit{random and statistical}: the secret is encoded within the statistical properties of a text, e.g., the 
distribution of word lengths, or the frequencies of vowels and consonant phonemes. Such techniques could
be of limited impact in smartphones, since they work efficiently when in presence of very large texts; 

\item \textit{linguistic}: the secret message is hidden within the structural flavor of the message, and could be used for instance in e-mails \cite{stemail} or in chats \cite{stechat}.  
\end{itemize}

\subsection{Methods Potentially Targeting Smartphones}

A notable group of techniques to be mentioned are those explicitly considering the features of MPEG data, which is the preferred format to deliver and store videos over smartphones. The two most important aspects of MPEG exploitable as steganographic carriers are the video’s I-frames’ color space \cite{mpeg} or P-frames’ and B-frames’ motion vectors \cite{video}.  However they were never tested in such devices. While also cost-effective phones are able to handle MPEG (also in a streaming manner), 
proper experiments would be needed to quantify if the hardware is able to extract the hidden data without impairing
the playback, or leading to excessive battery drains.  

For what concerns text, smartphones also support desktop-grade  content-rich document formats, such as Microsoft Word \cite{stegword} and PDF files \cite{stegpdf}, which can be used to hide information within their sophisticated data structures and metadata. In this case, tests to understand wether the mobile counterpart uses the very same formats and not limited capabilities implementations are needed. We underline that we put such methods here for prudence, since the 
everyday routine of cross-editing among phones and desktops do not hint at any differences in their formats.  

\subsection{Tools Implementing Object Methods}
\label{sta}

\begingroup
\renewcommand*{\arraystretch}{1.10}
\begin{table*}[htbf]
\caption{Steganography tools available for Android and iOS from application stores}
\begin{center}
\begin{tabular}{|c|c|c|}
\hline
\textbf{Type of Steganography} & \textbf{OS} & \textbf{Tools}\\
\hline
& & CryptApp, PixelKnot, Secret Letter, Shadow, Da Vinci Secret, \\   & & Camopic Lite, Steganography Application, Crypsis,   \\ Image & Android & MobiStego, Stegais, Stego Message, Limage, StegoLite,   \\ & & Secret Tidings, PtBox,  Steganografia, Incognito,  \\ & & Secure Message, VipSecret, Data Guardian\\
\cline {1-1} \cline{3-3}
Audio &  & CryptApp, StegDroid Alpha \\
\cline{1-1} \cline{3-3}
Video &  & CryptApp, Steganography Application \\ 
\cline{1-3}
& & iSteg, PicSecret, StegoSec, SpyPix, Hide It In, \\
Image & iOS &  Steganographia, Pixogram, Steganomatic, \\  & & Stega, Concealment, Invisiletter, iClic Reader \\ 
\hline
\end{tabular}
\end{center}
\label{tabtools}
\end{table*}
\endgroup

Here we present a concise analysis of smartphone applications enabling the use of information hiding techniques. We focus on tools available on official stores, therefore retrievable in an easy manner for the most popular devices. Besides, we solely investigate Android and iOS applications, since more than 90\% of current smartphones run such OSs \cite{idc}. 

On the average, applications offer basic functionalities, thus mainly resulting helpful if the user wants to conceal the secret information into a locally-stored carrier (usually, within a digital image). Such a carrier can be shared by means of communication services offered by the device (e.g., MMS, emails, or posted on a shared album over the Web). Obviously, the target must be aware of the steganographic procedure used, hence enabling the creation of an object covert channel. 

As today, there are a number of steganographic tools that are available via Google Play ($\sim$$20$) or Apple AppStore ($\sim$$10$). Table \ref{tabtools} provides an overview of the most popular steganographic tools. We underline that the majority shares some common features, such as: 

\begin{itemize}
\item many tools only rely on \textit{image steganography} also using basic techniques, resulting into easily discoverable channels. On the contrary, the most sophisticated and modern smartphone steganography methods are still not publicly available; 
 
\item about the totality of tools combines information hiding technique with some cryptographic algorithm to cipher the data before embedding the secret; 

\item a good amount of applications uses the multimedia features offered by modern smartphones. In this vein, it is possible to hide secrets in digital objects (e.g., images) already stored in the device, as well as to use the camera/microphone to produce carriers on-the-fly. 
\end{itemize}

Also, it is important to note that having such applications installed on the smartphone would be an effective indicator of steganographic attempts, eventually rising suspicious of an attacker (in the sense of a warden). Thus inspecting the activity of an user over a store (for instance, by analyzing whether he/she rated the application), or by scanning the device for local applications, could be used to void the steganographic conversation. Thus, all suspected carriers can be subjected to a detailed analysis making hidden data vulnerable. 

Similarly, if the steganographic tool has been removed (uninstalled in the \textit{jargon}) it is also possible to scan the device for related artifacts left out in the host OS, or to find signatures within files handled by known steganography methods. We point out that, this is how some commercially available steganalysis tools operate \cite{sarc}.

To summarize, applications presented in Table \ref{tabtools} are just simple proof of concepts to better understand internals of information hiding, rather than effective tools for production-quality environments.  In other words, they cannot be considered as definitive or effective for the clandestine exchange of confidential data.

\section{Smartphone Platform Methods}
\label{lcc}

\begin{centering}
\begin{figure}[t]
\center
  \includegraphics[width=7.5cm]{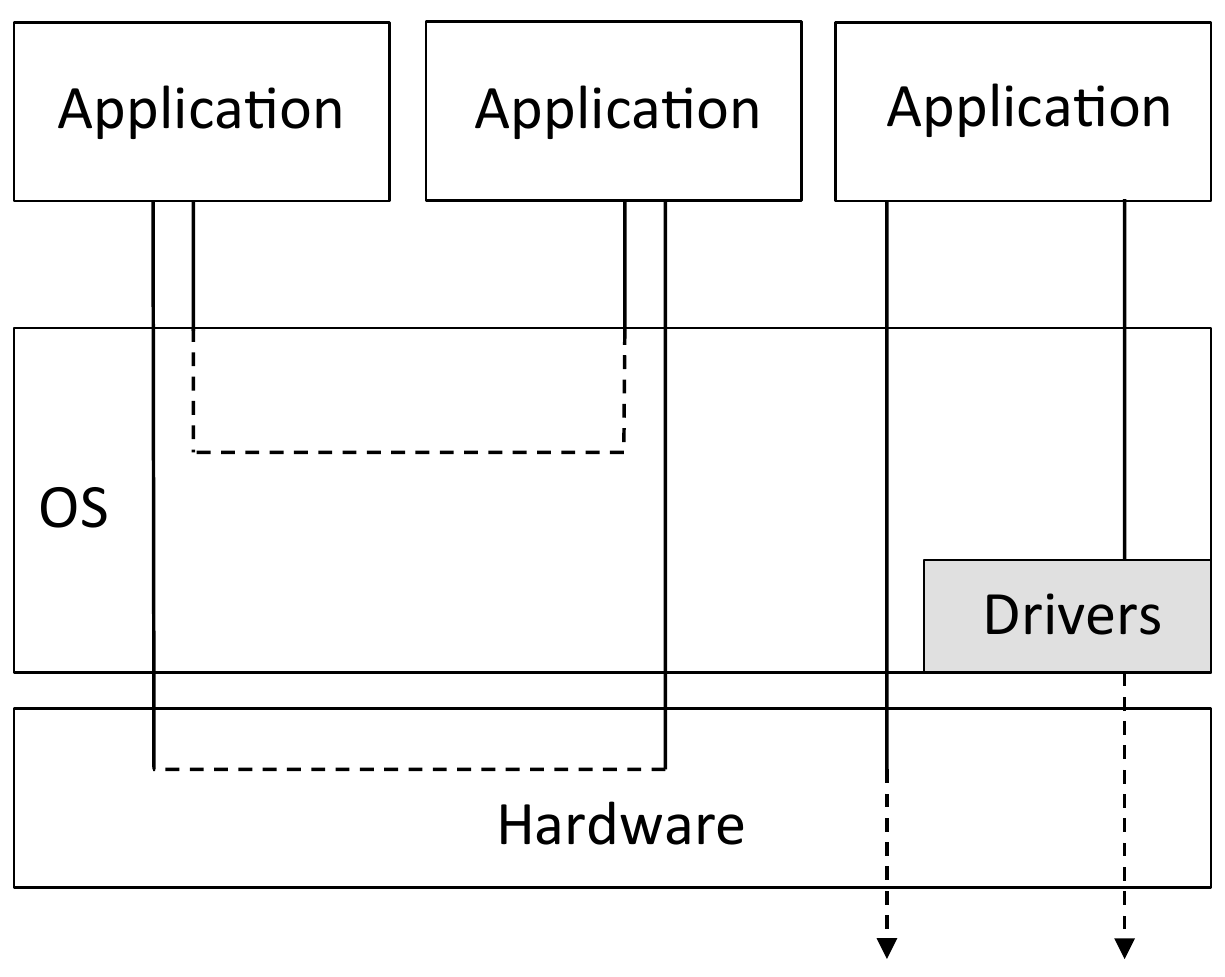}
  \caption{Different portions of a generic smartphone platform used to 
  implement hidden channels (in dashed lines in the figure). A local covert channel is created when the communication happens through two applications running on the same device. If the target is located outside the device, a network covert channel is needed.}
\label{stegplat}
\end{figure}
\end{centering}

As hinted, modern smartphones have an architectural blueprint very similar to the one of a standard desktop. Accordingly, their reference platform is composed of three major components enabling steganography: \textit{i}) the hardware,  \textit{ii}) the OS, and \textit{iii}) a collection of device drivers. 
In this section we concentrate on how they can be exploited to implement covert channels. As depicted in Figure \ref{stegplat}, a smartphone platform can be targeted for two different purposes:  
 \begin{itemize}
 \item to enable two (or more) applications running over the same device
 to communicate. In this case, a local covert channel is established, and to this aim, the main used components are \textit{i}) and \textit{ii});   
 \item to exfiltrate data outside the device, resulting into a network covert channel. For such a task, the main entities exploited are \textit{ii})
 and \textit{iii}). 
 \end{itemize}
 
Platform methods are primarily adopted to create a local covert channel and seldom employed to leak data to a remote party.  
 This is mainly due to the rich set of options available through the communication services offered by the device (discussed in Section \ref{ncc}). 
A notable exception is the case of malware, which use the OS as the preferred target to perform the majority of covert communications \cite{surwild}.  

Platform methods implementing a local covert channels are of paramount importance to bypass the framework enforcing security within a smartphone. Therefore, 
their evolution highly depends on how modern mobile OSs enforce applications, or third party libraries, to behave securely. 
Despite differences due to vendor-specific solutions, the standard mechanism is to use a \textit{sandbox} with well-defined privileges. Within such a
perimeter, the application can only access to some kind of resources (e.g., the address book),
or functionalities (e.g., network services). The creation of this protected environment 
varies according to architectural choices. For instance, Android devices use techniques
built on top of ad-hoc instances of the Dalvik Virtual Machine \cite{androidper}, while 
iOS relies upon a more layered approach \cite{iossand} (see reference \cite{confronto} for a \textit{vis-\`a-vis} comparison
of the different implementations). For the sake of completeness, other possible mechanisms are based on
hypervisors \cite{hyper}, or early detection of hazards by checking the application
 through the publishing pipeline from the developer to the ``marketplace" \cite{stores}. 

\begin{centering}
\begin{figure}[t]
  \center
  \includegraphics[width=8.5cm]{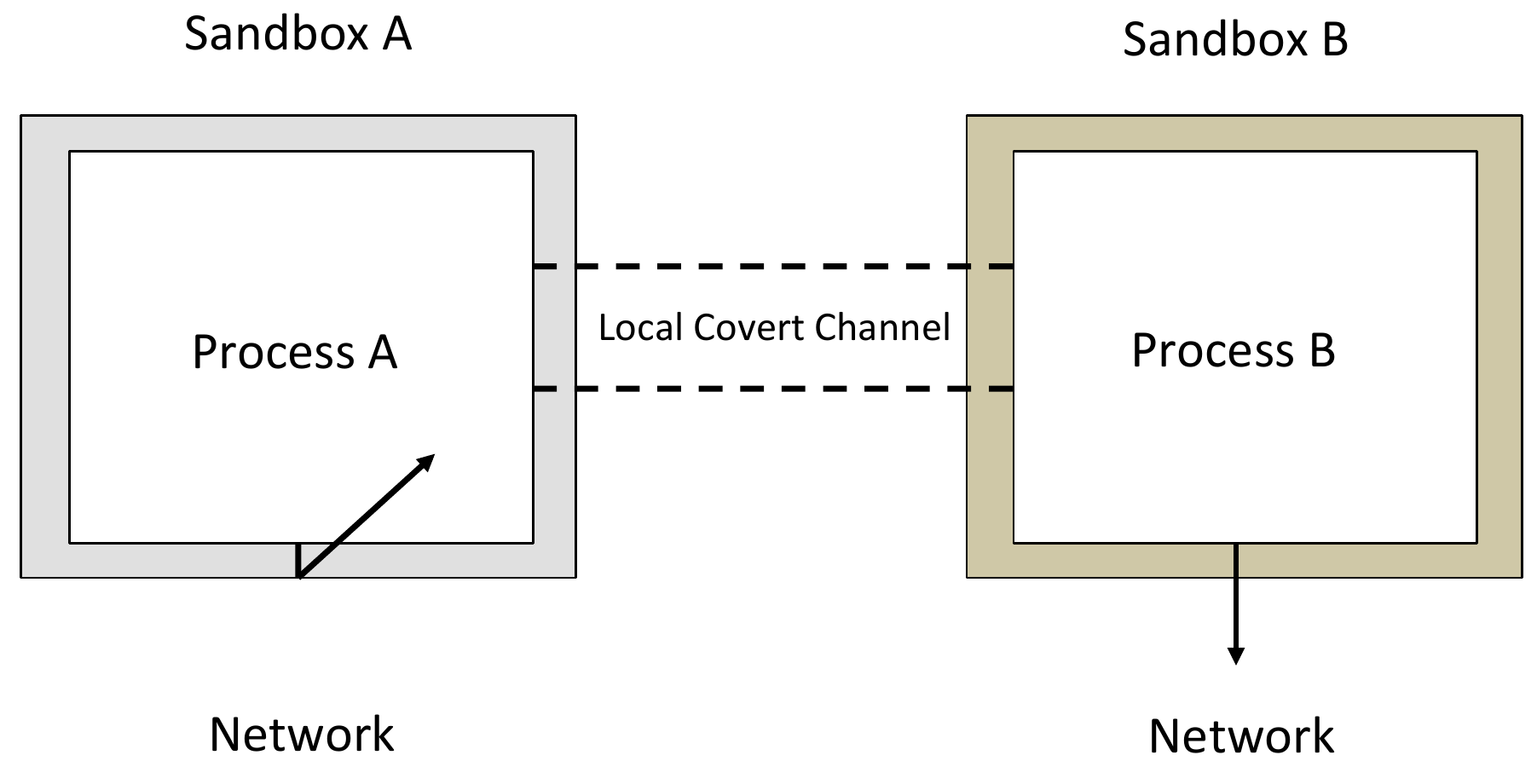}
  \caption{Reference usage of a local covert channel in a smartphone. {\tt Process A} wants to access the network, but {\tt Sandbox A} prevents all the communications. Therefore, it creates a local covert channel with the colluding {\tt Process B}, which has the proper privileges.}
  \label{localex}
\end{figure}
\end{centering}

Figure \ref{localex} depicts the quintessential usage of a local covert channel to bypass a sandbox. 
Let us consider two processes, {\tt Process A} and {\tt Process B}, running within {\tt Sandbox A} and
{\tt Sandbox B}, respectively.  Also, let us assume that {\tt Process A}  wants to use the network, e.g., 
to implement a C\&C channel to contact remote machines, but it has not the proper privileges. Then, steganography
is typical used as follows: 

\begin{enumerate}
\item {\tt Process A} is the actual malware, while {\tt Process B} is a companion application (in some papers they are defined as  \textit{colluding applications}). 
The latter is engineered to appear innocuous, and recipient of proper privileges. For instance, it could be an organizer tool needing the access to the Internet for remote syncing, which is a ``reasonable" requirement; 

\item {\tt Process A} communicates to {\tt Process B} through a local covert channel;

\item upon receiving data, {\tt Process B}  uses its privileges within {\tt Sandbox B} to leak information. 
\end{enumerate}

Obviously, communications of {\tt Process B} leaving the device could be detected, for instance by inspecting the traffic. 
Therefore, it is not uncommon to use another steganographic method to build a network covert channel (described in Section \ref{ncc}).  

\begin{centering}
\begin{figure}[t]
  \center
  \includegraphics[width=8.5cm]{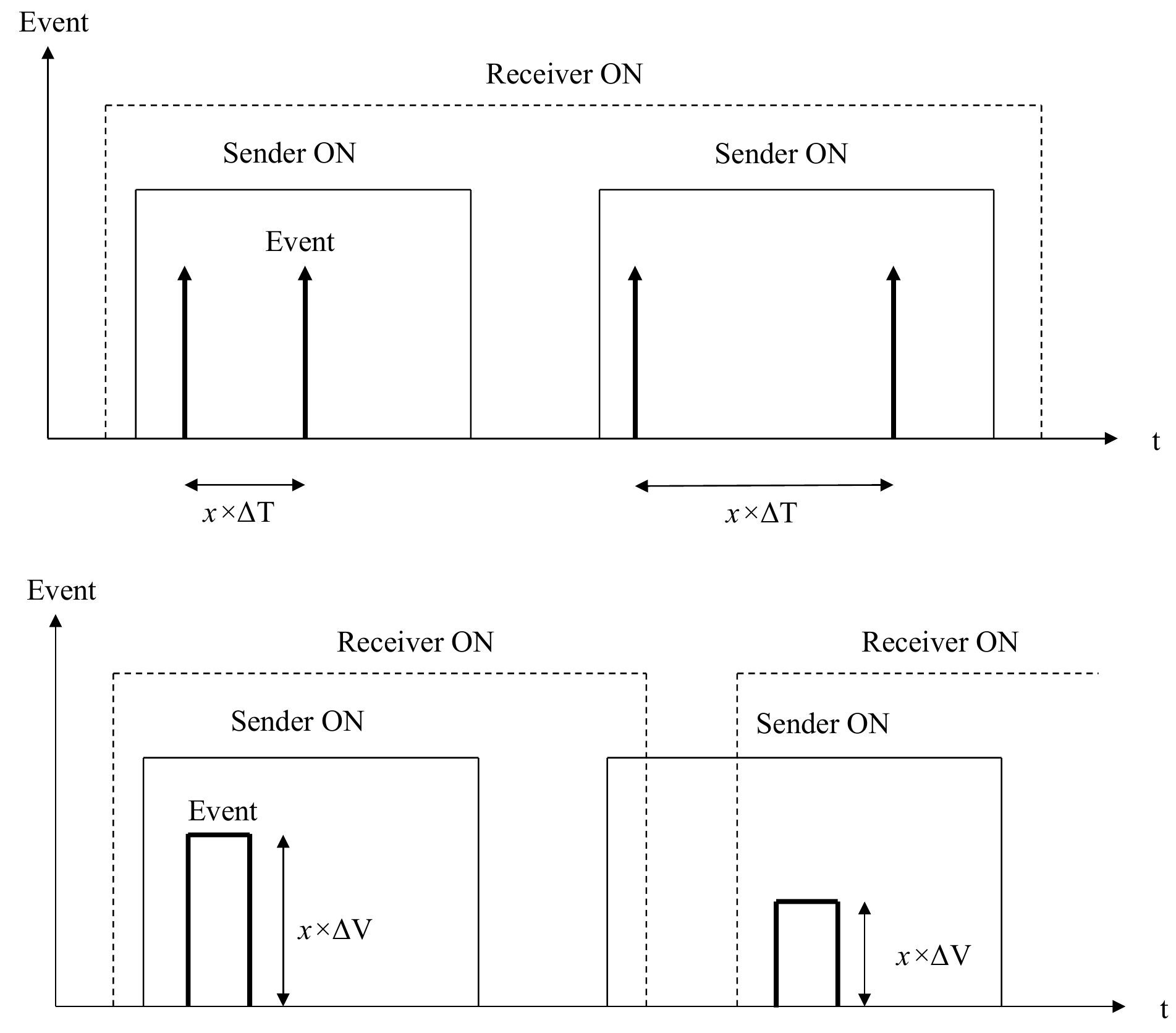}
  \caption{Reference models for establishing a local covert channel within modern smartphones. Upper: temporal. Lower: value.}
  \label{cchans}
\end{figure}
\end{centering}

Figure \ref{cchans} generalizes in more details the two main paradigms used to encode information 
to perform a data percolation among two processes, named \textit{sender} and \textit{receiver}. 
In the following, the general term \textit{event} is used to define a shared entity that can be manipulated by the sender and observed by the receiver, e.g., the state of the display, or the CPU load. In order to properly 
work, this mechanism requires that the sender and the receiver are both in a running state (labeled as
 {\tt ON} in Figure \ref{cchans}). However, the underlying birth/death evolutions ruling the processes can be partially decoupled (as depicted
 in the two examples of Figure \ref{cchans}). 

According to the literature, the current state of the art is to exploit \textit{two} main data hiding techniques: 
\begin{itemize}
\item \textit{temporal}: in this case the shared event is repeated in time. Therefore, 
the secret information is encoded within an integer value $x$ representing the 
time gap between two adjacent occurrences, i.e.,  $x \cdot \Delta T$. The 
main drawback of the approach is due to the lack of precision within the 
scheduler of the OS, thus limiting the achievable bandwidth of the resulting
steganographic channel; 

\item \textit{value}: the secret data $x$ is embedded within an alterable
value, e.g the CPU load, hence leading to a $x \cdot \Delta V$. This method is less 
prone to glitches due to the lack of real-time support of mobile OSs, yet it 
is fragile to noise altering the manipulated value. 
\end{itemize}

We point out that the previous approaches can be mixed as to enhance the bandwidth
of the covert channels, but by paying a cost in term of complexity, fragility, and reduced stealthiness.   

\subsubsection*{Example - Colluding applications via vibration settings} consider two applications enforced in their sandboxes wanting to exchange data. Each attempt will be
blocked, but they can be programmed in advance to create a hidden channel
(Figure \ref{localex}). For the sake of the example, let us suppose they can control the vibration setting of the device by using a specific system call. Therefore, they can use a portion of the OS to actually implement the channel (Figure \ref{stegplat}). For instance, a secret bit can be encoded by the sender application by setting/unsetting the vibration mode according to a specific pattern of temporal intervals (Figure \ref{cchans}). The target, by observing the status of the ringtone (i.e., enabled or disabled), can then reconstruct
the secret, therefore resulting in a data movement between the two processes, even if isolated. 


\subsection{OS-based Methods}
All of the existing OS-based covert channels have been proposed for Android operating system, 
probably mainly by virtue of its opensource nature. In fact, we did not found any paper dealing with  other mobile OSs (also going back to discontinued platforms like the Symbian).

The first paper that introduced local covert channels has been proposed by Schlegel et al. \cite{soundcomber}, and it uses information hiding techniques to implement a malware called Soundcomber. The latter is able to capture personal user data, e.g, the digits entered on the keypad of the device during a phone call to a customer care/bank call center. Soundcomber introduces \textit{four} different local covert channels, each one exploiting a well-defined functionality of the smartphone. 

In more details: 

\begin{itemize}
\item \textit{vibration settings}: the secret data is exchanged between two colluding applications by differencing the status of the vibration, i.e., the silent mode is toggled. One application modifies the settings and the other interprets this switch as binary data. The achieved steganographic bandwidth is $87$ bps, and it has been evaluated through experiments on a real prototype;

\item \textit{volume settings}: the secret data is exchanged between two colluding applications by selecting different values for the volume of the ringtone. In principle, the covert channel is similar to the previous one, but the availability of more values
(different sound intensities versus vibration has only on/off states) enables to experimentally achieve a steganographic bandwidth of $150$ bps;

\item \textit{screen state}: the secret data is exchanged between two colluding applications by acquiring and releasing the \textit{wake-lock} permission that controls the screen state. If such lock is held for a sufficient time, then the screen is prevented from turning on. Otherwise, the OS sends a notification. This process can be properly modulated by the sender, and trials demonstrate that the resulting capacity is $\sim$$5.3$ bps;
\item \textit{file locks}: the secret data is exchanged between two colluding applications by competing for a \textit{file lock}. If one application wants to send the bit {\tt 1}, then it requests a file lock on a commonly shared file. If also the receiver tries to lock the same file, two outcomes would be possible: the bit {\tt 1}, if this attempt fails, or the bit {\tt 0} otherwise. Manipulating such behavior leads to a steganographic bandwidth of $685$ bps, according to experiments done by the authors. 
\end{itemize} 

The work by Marforio et al. \cite{collapp} extends such ideas, by exhaustively investigating the usage of different locks simultaneously, mainly with aim of improving the steganographic capacity of the covert channel. They propose \textit{seven} 
different variations, and also provide a performance evaluation of prototype implementations by using Android smartphones. Specifically: 

\begin{itemize}
\item \textit{single and multiple settings}: similarly to Soundcomber, the covert channel is created by concomitant alterations of the vibration/volume settings. In this manner, the steganographic bandwidth 
can be expanded up to $\sim$$260$ bps, which is an increase compared to the $87$ bps and $150$ bps achieved in \cite{soundcomber} via the disjoint manipulation of  vibration and volume settings, respectively; 

\item \textit{automatic intents}: in the Android OS, an intent is an asynchronous object message allowing an application to request functionality 
from other components (e.g., launch the address book) \cite{intent}. In this case, the secret data is exchanged between two colluding applications by using OS-wide notifications as the carrier (e.g., the address book has been opened, or the address book has been locked). Specifically, the method exploits broadcast messages automatically sent to interested processes  when the modification of a commonly subscribed setting occurs (e.g., all the applications are notified that the address book has been changed). The resulting experimentally achieved steganographic bandwidth is $\sim$$71$ bps;

\item \textit{type of intents}: the secret information is exchanged between two colluding applications by encoding data into a type of intents, which is the kind of the event of interest, rather than its value. Possible examples are, subscribing for 
changes into a set of flags. Then, the percolation is done in a similar way of automatic intents. Since an intent is a form interprocess
communication, its utilization should be carefully evaluated, as to avoid
additional security risks besides steganography \cite{intent}. This approach yields a quite high steganographic bandwidth of $3,837$ bps;

\item \textit{threads enumeration}: the covert channel is created between the two colluding applications by encoding data into the number of currently active threads reported into the {\tt /proc} directory. The achieved experimentally steganographic bandwidth is $148$ bps, on the average; 

\item \textit{Unix socket discovery}: similar to the threads enumeration, but the secret is encoded into the state of a socket. The resulting steganographic bandwidth is $2,129$ bps;

\item \textit{free space on filesystem}: the hidden transmission happens by encoding data into the amount of free space in the storage unit of the device. The number of free blocks is controlled by the transmitting side by performing ad-hoc crafted {\tt write/delete} operations. The resulting steganographic bandwidth is $11$ bps;

\item \textit{reading the} {\tt /proc/stat}: the secret data is exchanged by encoding the information into the processor usage statistics available in {\tt /proc/stat}. For instance, the transmitting side intentionally performs some computations to artificially alter the system load, while the receiving side is monitoring the CPU usage, as a a way to infer the transmitted secret data bits. Trials support a covert channel capacity of $5$ bps.
\end{itemize}

In the recent work by Lalande and Wendzel \cite{wendzel} additional \textit{four} ways to enable local covert channels are discussed. However, according to the used taxonomy, only three are OS-based, thus one will be presented in Section \ref{hbm}. 
While designing the channels, authors primarily focused on how to reduce their ``attention raising" characteristics. But, as a consequence of the magic triangle rule, this will lead to scarce steganographic bandwidths. In details, the proposed methods rely upon: 

\begin{itemize}
\item \textit{task list and screen state}: the hidden exchange between two colluding applications is realized by encoding secret data bits into the time that the sender application stays active after the screen is switched off. Based on this measurement, and under assumption that an user will not interrupt the receiving application, the latter can infer secret data bits value. The achieved steganographic bandwidth, which was only evaluated experimentally, is very limited: $\sim$$0.3$ bps; 

\item \textit{process priority and screen state}: this method utilizes a synchronization scheme based on screen off states in a way similar to the previous one. But, in this case, the receiving side measures the time during which a given process priority remains set to a reference value. The sameness also reflects in the  attained bandwidth of $\sim$$0.3$ bps; 

\item \textit{process priority}: this is a simplified version of the former technique, since the synchronization part is omitted. The receiver scans the priorities of all possible processes until it detects the predetermined value. Then, the secret data is sent in a similar manner. Such approach greatly improves the steganographic bandwidth up to $17.6$ bps.
\end{itemize}

Notably, the work by Lalande and Wendzel \cite{wendzel} also estimates how their methods would impact on the energy consumption of the smartphone. This paper is an exception, since the literature dealing with smartphones and steganography almost neglect this topic. Although, battery drains due to hidden communication should be always explicitly addressed, since they can be exploited for detection purposes. 

Their experimental results \cite{wendzel} demonstrate that the \textit{process priority and screen state} method is approximatively two times more efficient in terms of energy consumption compared to the \textit{task list and screen state} and the \textit{process priority}. Besides, the energetic footprints for all the methods are almost masked by the consumption of the kernel used by Android, which is $80$ mJ/minute. However, the local covert channel exploiting \textit{process priority} is very 
energy consuming on the long run, since the receiver needs to scan each time a large number of Processes IDentifiers (PIDs) in order to discover the process that will change its priority. This can lead to a quick depletion of the battery of the hosting smartphone, potentially revealing the steganographic attempt. Notice how this conforms again with the magic triangle rule. 

\subsection{Hardware-based Methods}
\label{hbm}
Compared to previous methods, hardware-based ones act on the very low levels of the device, and primarily concentrate to create an OS-independent local covert channel. As today, this technique is still not very popular in the smartphone panorama. 

Marforio et al. \cite{collapp} engineered \textit{two} local covert channels for Android-based smartphones relying on processor statistics:
\begin{itemize}
\item \textit{system load}: to encode secret bits, two patterns are used (i.e, the 
$x \cdot \Delta V$ depicted in Figure \ref{cchans}). The bit {\tt 1} is represented by the sender performing a variety of computationally intensive operations, while the bot {\tt 0} is coded by switching the process to the idle state (in other words, the sender relinquishes the CPU). During experiments, the reached steganographic bandwidth peak is $3.7$ bps;

\item \textit{usage trend}: as a variation of the previous mechanism, here the receiving side infers data by monitoring the trend of the processor frequency by repeatedly querying a proper system-wide interface. This method is more efficient, and  accomplished a steganographic capacity of $\sim$$4.9$ bps.
\end{itemize}

Also, the work of reference \cite{wendzel} contains a variation of the \textit{screen state}, but relying on the behavior of the hardware. In this case, the sender switches the screen on and off to send secrets. To decode the hidden information, the receiver measures  the time elapsed between two consecutive screen active states (i.e., the 
$x \cdot \Delta T$ depicted in Figure \ref{cchans}). The resulting steganographic bandwidth is $\sim$$0.22$ bps.

For what concerns other hardware components of devices, Cheong et al. \cite{nfcsteg} exploit both NFC and steganography to use a smartphone
as an access control system, also with the scope of increasing the security of de-facto standard
contact-less protocols. In details, the NFC is used to interact with elements of a building, while
the access keys needed to actuate commands (e.g., unlocking doors) are exchanged via 
hidden data embedded within the user profile picture. 

Another example of internal hardware 
used for steganography is given by \cite{accelsteg}, where a method using accelerometers to measure self-induced or external 
vibrations to establish a covert channel is explained.  

Lastly, the work of Wang et al. 
\cite{sdsteg} deals with a technique 
exploiting removable flash memory cards, which are ubiquitously available
in many devices. In details, authors propose to use the
RC4 algorithm to choose the locations of the flash where to hide secret bits. Then, each one is encoded by altering the electrical characteristics of the card, by means of a pattern of ``stress cycles", which are the number of times a single bit is written (i.e., set from {\tt 0} to {\tt 1}), or erased (i.e., set from {\tt 1} to {\tt 0}). The hidden data can be shared among two local processes to exploit 
a local covert channel, or upon removing the flash memory, physically moved towards the interested target (as an ``object").

 \subsection{Methods Potentially Targeting Smartphones}
 \label{mpts2}
 
A very active field of research in steganography applied to hardware is related to 
the exploitation of internals of the IEEE 802.11, which is an ubiquitous standard
for wireless connectivity for smartphones and mobile appliances. Even if such 
techniques aim at sending data outside the device (according to the taxonomy
presented in Section \ref{tsts}), we review them here, since they
heavily rely on the interaction with low-level components of air interfaces, 
such as the hardware responsible of performing framing or modulation. 
Besides, such techniques particularly require modifications within the
device driver, thus completing the review of steganographic channels exploitable through the core elements implementing a smartphone. 

In more details, the components of the IEEE 802.11 architecture that can be used are:  
\begin{itemize}
\item Orthogonal Frequency-Division Multiplexing (OFDM) within the IEEE 802.11n:
to strengthen the signal against obstacles and reflections, the IEEE 802.11n imposes the 
usage of a guard interval among two transmissions, which is filled with a cyclic prefix. 
The latter improves the quality of the transmission since it affects the radio 
propagation, but its content is discarded by the receiver. The method proposed in \cite{ofdm}
embeds data within the cyclic prefix, and its performance in terms of steganographic 
bandwidth varies according to the used modulation:  $3.25$ Mbps in Binary Phase Shifting
Keying (BPSK), $6.5$ Mbps in Quadrature Phase Shifting Keying (QPSK),
$13.0$ Mbps in 16-Quadrature Amplitude Modulation (QAM), and $19.5$ Mbps in 64-QAM; 

\item Frame Control Field (FCF) within the IEEE 802.11a:  the work proposed in \cite{80211steg}
analyzes different fields of the FCF used in the Medium Access Control (MAC) header of 
IEEE 802.11 Protocol Data Units (PDUs). Authors conclude that the most transparent fields are the {\tt Retry}, 
{\tt PwrMgt} and {\tt More Data}, enabling the transmission of hidden messages with a rate of $16.8$ bps. 
Even if the method is simply detectable, it can be easily implemented within the device driver of many
Linux-based smartphones (i.e., Android);

\item bits used for padding frames in the IEEE 802.11g: the method discussed by Szczypiorski and Mazurczyk \cite{wipad} 
inserts secret bits into the padding of transmission symbols. Authors also propose how to 
enhance its steganographic capacity by using two separate covert channels, i.e., one living within the padding of data frames having a capacity of $1.1$ Mbps, and another one within ACK enabling to convey secrets at $0.44$ Mbps (therefore, $\sim$$1.65$ Mbps when combined); 

\item hiding data in frames forged to be corrupted: the method presented by Szczypiorski \cite{hiccups} proposes
to waste a percentage of the overall transmission bandwidth by producing corrupt frames containing
hidden data. However, an anomalous Frame Error Rate (FER) could reveal the steganographic communications. Thus, author proposes to use a FER of $2.5$\%, which is normal considering average environments and mobility, and evaluate that in an IEEE 802.11g WLAN this method can achieve a channel capacity of $216$ kbps.  
\end{itemize}

Unfortunately, such techniques need a relevant amount of tests and of engineering to
be successfully ported on smartphones. In this vein, the major research items needed
to bring them ``on the road" are: 
\begin{itemize}
\item they require the access to device drivers: while this 
could be trivial for Android, it is very strict precondition for other platforms. In addition, 
device/kernel-level manipulations may lead to system instability, then 
such methods must be carefully architected;  

\item they manipulate the physical/data link layers of the ISO/OSI stack:
despite the availability of drivers, some low-level features could be 
non accessible by the software. Therefore, proper workaround 
must be made. Nevertheless, the heterogeneity of features offered
by the implementations ``on the wild" (e.g., stacks/interfaces having reduced capabilities)
should be properly evaluated with a per-device granularity, hence
resulting into a very complex problem space; 

\item they use a relevant amount of power: wireless transmission, as well as spinlock-like code needed to intercept data within the stack, are both energy and CPU consuming tasks. Therefore, this kind of covert channels could impair the smartphone, or cause unacceptable battery drains. 
\end{itemize}

\section{Smartphone Communication Methods}
\label{ncc}
Since smartphones have a full-featured TCP/IP protocol stack, it is possible to theoretically exploit the majority of the network steganography methods already proposed for desktops (see, \cite{survey2}, \cite{survey1} and \cite{bender} for surveys on the topic), for instance, those that are based on packet timing or header manipulation. Our review reveals that novel network covert channels specifically crafted for smartphones are still a small niche. 
Notwithstanding, they are seldom used as standalone tools, since they make useful local covert channels discussed in Section \ref{lcc}. In fact, many of the works dealing with
colluding applications (e.g., \cite{soundcomber}, \cite{collapp} and \cite{wendzel}) have the preliminary goal of exfiltrate data to a process having proper 
network privileges (for instance, the {\tt Process B} depicted in Figure \ref{localex}). 

Then, we firstly survey channels dealing with legacy services, and then those 
using smart communication capabilities. 

\subsection{Legacy Services}
\label{lst}
With the acceptation of ``legacy services" we define functionalities
related to the telephonic portion of the smartphone, such as SMS and MMS,
which are still very relevant carriers for steganography.  
In fact, according 
to \cite{gsm}, in the second quarter of 2007, Verizon Wireless alone delivered $28.4$ billion of text messages. 

Before introducing techniques, we quickly review SMS and MMS to 
give the proper foundation. Put briefly, SMS enables to send up-to $160$ characters encoded into
a $140$ byte-long payload. The number of characters varies according 
to the alphabet used by the subscriber, e.g., $160$ characters when $7$ bit encoding is used, 
$140$ or $70$, when using $8$ or $16$ bit encoding, respectively. We point out
that the SMS service can be also used to transmit very basic 
$72 \times 28$ pixels B/W pictures.  
Many SMS can be concatenated to bypass such limit, thus allowing a maximum of 
$255$ chained messages. However, many carriers limit the number of subsequent segments
to $8$. 

Concerning the MMS services, it is more complex, since it supports
different data, ranging from voice to audio, hence requiring proper 
adaptation layers and interfaces \cite{ppmms}. Similarly to SMS, 
also MMS has a narrow capacity: text is limited to $30$ kbyte, images to $100$ kbyte, and
multimedia contents must be smaller than $300$ kbyte.

References \cite{sms11} and  \cite{sms12} exploit SMS to hide 
data within a low resolution picture. The latter is subdivided in blocks
of $3 \times 3$ pixels, and each one can be set to $1$ (black) or $0$ (white). 
Therefore, each block leads to $2^9=512$ possible ``shapes". Within a block,
some pixels can be flipped to embed secrets: even if this introduces noise, 
a limited amount of changes could not be noticed (i.e., authors evaluate a 
maximum of $27$ bytes per image before the method becomes useless). At the same time, the basic nature
of two-color ultra-low resolution pictures has the following additional drawbacks: 
the lack of a richer grayscale makes more difficult to hide data, and 
this carrier has a very low degree of resistance against noise. 

The works available in references \cite{sms1} and  \cite{sms2} are more focused in the
textual properties of the SMS. Specifically, \textit{three} base methods, which can be 
mixed to enhance the throughput of secret data, are proposed: 
\begin{itemize}
\item \textit{acronym patterns}: words are substituted by means of acronyms (e.g., 
``CM" stands for ``call me", or ``C" stands for ``see") to build a binary pattern of 
{\tt 1} if the acronym is used, and {\tt 0} otherwise, as to embed secret data; 

\item \textit{alternate spelling}: similar to the previous one, but uses known alternate
spellings of a words (e.g., ``center" instead of ``centre"). A similar technique is based
on using synonyms;

\item \textit{white spaces}: the secret data is represented via well-known patterns of spaces among
the words, (e.g., {\tt Hi\textvisiblespace\textvisiblespace how\textvisiblespace are\textvisiblespace\textvisiblespace you?}). An equivalent method uses line skipping. 
\end{itemize}

Shirali-Shahreza et al. \cite{sms13} developed a method based on Sudoku puzzles embedded within the SMS is presented. 
In details, information is hidden within the permutations of a $9 \times 9$ Sudoku scheme. 

Rafat \cite{sms3} exploits the availability of two standard font classes, namely proportional
and system, as a way to produce secret patterns within the SMS via ad-hoc font changes. 

For what concerns the steganographic capacity, all of the surveyed works do not report a detailed analysis of the
achievable throughput. But, the extremely simple nature of SMS communication always reflect in a very limited bandwidth,
which could be slightly increased by sending concatenated messages. Besides, when in 
presence of embedded images, their B/W nature, jointly with the very limited 
number of pixels available rise the same bandwidth problems, and also make
the steganogram easily detectable, and prone to noise, thus making such methods
very fragile. 

As regards MMS, references \cite{mms1} and \cite{mms2} compare different steganographic methods, especially crafted for this carrier. Summarizing: 

\begin{itemize}
\item \textit{LSB encoding}: the message is stored within the LSB of each pixel composing the image sent via MMS. On the average, the LSB could be assumed as random noise, thus its alteration would not have visible effect on the image. Possible variations include multiple bits changing, or adding pseudo noise to make the detection harder; 

\item \textit{echo hiding}: the information is injected within the audio of the MMS, by adding an echo generated via a proper kernel and modulated according to the data to be hidden, also considering the limited nature of this carrier; 

\item \textit{markup embedding}: an MMS can embed also portion of data described through mark-up languages, such as the  eXtensible Markup Language (XML), HyperText Markup Language (HTML) or Synchronized Multimedia Integration Language (SMIL). Then, tags can be reordered to hide data, since they do not cause any changes to the rendered text. Unneeded tag, which are usually ignored by the parser, can be exploited for information hiding purposes. Besides, since spaces and white lines are ignored as well, the source code of the markup can be 
manipulated via the aforementioned white spaces/line skipping methods. 
\end{itemize}

Mohanapriya \cite{mms3} introduces a technique using the manipulation of the array of coefficients of the DCT used in the Joint Photographic Experts Group (JPEG) format. Similarly, Singh et al. \cite{mms4} use a low-power signal generated via a spread spectrum modulation within an image. In details, a Pseudo Random Noise (PRN) pattern is added to the carrier image and the seed used for its generation encapsulates the hidden data.  Owing to its reduced power, the hidden message cannot be detected both by humans or automated tools not having access to the original/unmodified data. This method is also more robust to the previous ones using the alteration of the LSB and DCT. 

Jagdale et al. \cite{mms5} discuss a solution to embed secret messages within a generic carrier (including MMS), by means of the aforementioned techniques. The main differences from the other methods is the prior encryption of the steganographic message by means of elliptic curve cryptography.

Similarly to SMS, also papers dealing with MMS do not contain an accurate estimation of
the steganographic bandwidth. However, since they mainly rely upon the same techniques used for
multimedia objects, but with reduced sizes, their resulting capacity is usually very limited. 
Therefore, both SMS and MMS appear as suitable only for very short covert conversation, or
as a method to send a very minimal set of information (e.g. a secret key) to unlock/decode
data delivered through a more efficient covert channel.

For what concerns voice, Miao et al. \cite{3Gsteg} exploit features of the Adaptive Multi-Rate Wideband (AMR-WB) encoder adopted in 3G communications. The approach highjacks the search phase of words belonging to the algebraic codebook used during the encoding process, i.e., some bits of the resulting codebook
contains the hidden data. Yet, a too aggressive alteration of codewords worsen the Signal to Noise Ratio (SNR), eventually resulting into detectable artifacts. Besides, the steganographic method must 
also take into account channel conditions, since the codec alters the features of its stream 
according to the measured SNR. For such reasons, authors proposed a time-varying
embedding factor $\eta$ used to control the trade-off among the SNR and  
steganographic bandwidth. The best tradeoff is achieved when $\eta=4$ leading to 
a steganographic bandwidth of $1,400$ bps, which can be used only 
for short periods and under error-free channels to avoid a quick degradation of the voice quality.  

A similar approach is presented in the the work presented in \cite{3gppsteg}, where the 
the 3rd Generation Partnership Project (3GPP) single-channel narrowband Adaptive Multi-Rate (AMR) codec is used. Also in this case, 
some bits of the codebook are altered resulting into a manipulation of both the frequency and
time envelopes of the voice signals. The resulting steganographic bandwidth is of $2$ kbps.  

As regards low-rate standards, Tingting and Zhen \cite{gcod} deal with the G.723.1 low-rate codec, offering a steganographic capacity of $133.3$ bps (i.e., altering the $5$ least significant bits of an encoded frame). 

Lastly, there are also many techniques using the characteristic of the codec used by 
legacy GSM services, which are outside the scope of this survey. For the sake of 
completeness, we mention, among the others, the work available in 
\cite{steggsm} that uses low power tones artificially added within the GSM-encoded
voice.
  
\subsection{Smart Services}
As presented, with the acceptation of \textit{smart services} we defined all
the functionalities expanding the classical connectivity of the so-called
\textit{feature phones}. As an example of smart services, we mention 
Internet connectivity, and all the applications that can be used on top
of the TCP/IP protocol stack, such as Web browsers or multimedia streaming. 

One of the first network covert channel for smartphones was proposed by Schlegel et al. \cite{soundcomber}, and it is based on the utilization of a Web browser, where the request to open URL was intentionally formed as {\tt http://target?number=N},  where {\tt N} is substituted with the secret data string to be exfiltrated to the target website.

Gasior and Yang \cite{gasior} proposed two ideas to implement a network covert channel. In the first one, authors developed an application transmitting live video from the camera of an Android device to a remote server. Then, the steganographic method creates a covert channel by altering the delays between the video frames sent to the server. The experimental evaluation was performed in \cite{gasior2} and the resulting steganographic bandwidth is $\sim$$8$ bps.
The second covert channel used a tool that displays an advertisement banner at the bottom of a running  application. The content of the banner was selected from \textit{N} different advertisements available, and then requested to the remote server. The application leaks information by requesting a specific advertisement to represent a binary values of the sensitive user data, e.g. the contact list. 

Nowadays, it is surely true that OSNs are largely accessed from  smartphones
and mobile devices, then we address in more details steganographic methods
that conjugate social platforms and mobile devices. 
Castiglione et al. \cite{stegfoto} propose different techniques exploiting photos and
websites typically used by smartphone users for sharing, such as OSNs and 
Flickr-like destinations. In more details, authors propose to exploit features easily
manipulable from devices, namely: \textit{i}) altering the built-in naming convention
often used by devices (e.g., {\tt IMG\_xxxx.jpg}) to upload files having 
data embed in filenames, and \textit{ii}) exploit the feature named as ``tagging" (i.e., 
the possibility of annotating digital images by pinpointing names, locations or
GPS positions) to produce sequence of tags containing secret data. Tagging
is also widely supported by many ad-hoc client interfaces offering an access to 
OSN without needing the Web browser. Even if such methods are quite simple, 
they have the major benefit of not embedding information within pictures, thus
not requiring specific software running on the phone. Moreover, the procedure
resists against possible manipulation implemented by the remote site (e.g., 
resizing or recompression). 

In \cite{xmpp} the Extensible Messaging and Presence Protocol (XMPP), which is
at the basis of many OSNs like Facebook or Google+, as well as 
many Instant Messaging (IM) services also available for smartphones, like Google Talk and LiveJournal Talk
is discussed. In this case, the steganographic channel is created by 
encoding data within attributes exchanged by the protocol. For instance, the {\tt type} attribute,
which is case-insensitive, can embed secret data as follows:  \\{\tt type=`CHAT'}, {\tt type=`chat'} or {\tt type=`cHaT'}.

From another perspective, also the work by Kartaltepe et al. \cite{bottwi} portraits 
how the Twitter OSN could be used for steganography, even to coordinate
a botnet, or to implement C\&C functionalities.  
Specifically, a Twitter feeds can be populated with innocent looking messages (i.e., the tweets)
embedding secret command (for instance by using text-based steganography
as discussed in Section \ref{eds}). Thus bots can simply retrieve data by 
issuing an HTTP {\tt GET} to a public Twitter account. Commands can be posted 
from a smartphone, e.g., via the Twitter client-interface, or the mobile version
of the site. 

\subsection{Methods Potentially Targeting Smartphones}

As discussed, the availability of many desktop-grade applications multiply, 
at least theoretically, the number of network covert channels to be used
on smartphones. To mention the most popular applications used on 
mobile devices, we cite Skype, generic VoIP client interfaces, 
Facebook (see, e.g., the popular tool discussed in reference \cite{stegobot}) or the BitTorrent file-sharing service. Therefore, 
as desktops and smartphones progressively converge into a unique platform, we review
here those applications that, at the time of writing, or in near future, could be
used on portable appliances for steganographic communications. 
  
For the case of Skype, in 2013 a method named  Skype Hide (SkyDe) has been proposed by Mazurczyk et al. \cite{skyde}. It relies on a feature characterizing the traffic of Skype, since the actual implementation does not utilize any silence suppression mechanism. Consequently, PDUs carrying silence can be easily recognized despite encryption, as they are about half of the size of  PDUs containing voice.  Thus, talk/silence patterns embed the secret communication. The method tested in a real environment is able to provide a steganographic bandwidth of $\sim$$2$ kbps.  

For generic VoIP services, many different methods have been also proposed (refer to reference \cite{voipsurvey} for a very recent survey on this topic). For instance, we mention Transcoding Steganography (TranSteg) \cite{transteg} in which the overt data is compressed to free space exploitable for secrets. Put briefly, for a
voice stream, TranSteg finds a codec that will result in a similar voice quality but 
with a smaller payload for each PDUs. By transcoding the data flow without raising the suspicions of end users, the obtained additional free space can be used as a network covert channel. Experimental results show that TranSteg can achieve $32$ kbps when the G.711 codec is used. 

Another possible mechanism is implemented by the Lost Audio Packets Steganography (LACK) method \cite{lack1}, \cite{lack2}. LACK takes advantage of excessively delayed voice packets, which are usually discarded by many multimedia communication protocols. In this vein, such PDUs are highjacked, and a receiver aware of the steganographic exchange can extract the secret data. As pointed out, the performances in terms of capacity are tightly ruled by the codec. For instance, when using the G.711 codec the steganographic bandwidth is $\sim$$600$ bps. 

As regards popular p2p file-sharing platforms, we mention the work by Kopiczko et al. \cite{stegtorrent}, where authors propose a steganographic method for BitTorrent, named StegTorrent. The basic idea is to modify the order of the PDUs exchanged via the peer data exchange protocol (i.e., the one used for transfer data within a \textit{swarm}). The resulting steganographic bandwidth for this method is $\sim$$270$ bps.

Even if the aforementioned applications are already available on smartphones, their correct exploitation for the creation of a network covert channel still needs to be proven. Reasons are similar to those discussed in Section \ref{mpts2}, and they are omitted here for the sake of brevity.  

Lastly, we showcase another important class of applications that are already successfully utilized on desktops for steganography: network games. In fact, modern devices have 
enough power to run full-featured/desktop-class games, and their enhanced connectivity makes them a very
powerful platform for online gaming. An effective approach, which could be easily borrowed from desktops, is based on embedding secret data in the traffic 
used by First Person Shooters (FPSs) to sync the client(s) and the server.

In details, 
Zander et al. \cite{q3a} create a covert channel  by using the popular Quake III Arena FPS. To this aim, an instrumented client manipulates some bits of the angles used to locate the orientation of a player within the game world (also called the \textit{map}). Yet, the number of alterations should be minimal, since they add noises, lags or prevent players to correctly move around the map.
 Authors report that such technique offers performances in the range of $8-18$ bps. 
  
Similar ideas have been also used for board games, as reported in \cite{stegogo} and \cite{stecoll}.

\section{Mitigation Techniques}
\label{co}

As said, steganography leads to multifaceted privacy and security implications, thus a part of the research is devoted to reveal or prevent covert communications. Generally, steganography mitigation techniques aim at: \textit{i}) detecting, \textit{ii}) eliminating, and \textit{iii}) limiting the steganographic bandwidth (or capacity) \cite{survey1} of a covert channel. Especially,
for \textit{i}) there is a well-defined research area called \textit{steganalysis}, while for \textit{ii}) and \textit{iii}) the lack of universal solutions stems countermeasures tightly coupled with the specific information hiding technique. 
Therefore, in the following, the proper countermeasures for each type of covert channel are addressed separately.

\subsection{Countermeasures Against Object Covert Channels}

Mitigation of object covert channels is the most mature field among all the possible countermeasures methods. Besides, since modern smartphones can handle contents without limitations, the reference literature is almost the same for other computing devices. For such
reasons, we adhere on the classifications for detecting techniques developed in this field during the years. Even if they differ in some details, all of them partition \textit{steganalysis methods} into \textit{two} main branches \cite{steclass}: 
\begin{itemize}
\item \textit{signature}: the alterations of the media storing the secret information are detected via searching for known and well-defined patterns or features. For instance, for the case of {\tt .gif} images, their related color palette is checked against a database of well-known alterations  \cite{gifsteg};

\item \textit{statistical}: the hidden data is detected by inspecting how the manipulated media deviates from 
an average-known behavior. As a paradigmatic case, the alteration of the LSB within an image leads to 
a noise greater than the average for a reference class \cite{steansur} (i.e., \textit{threshold} detection). 
\end{itemize}

We underline that both mechanisms depend on the type of digital objects under investigation, even if the development of an ``universal" technique is still an open research topic \cite{steclass}. 

Once the covert communication is detected (i.e., the media embedding data is recognized) the  steganographic 
channel is eliminated by removing the hidden information from the digital object possibly without disrupting the legitimate communication. Such a process is also defined as \textit{sanitization} \cite{httpsan}. 

In the following, we review the main methods used for each type of object, as presented in Section \ref{occ}. 

\subsubsection{Images}
Ker \cite{lsban} deals with the detection of LSB manipulation in digital pictures by evaluating the Histogram Characteristic Function (HFC) of the image (i.e., the energy distribution of the image histogram in the frequency domain). HFC methods have been also successfully used to detect data embedded in images when in presence of more sophisticated procedures, like the spread spectrum, and the DCT \cite{lsban2}.  

Moreover, Fridrich et al. \cite{lsban3} introduce another method for detecting LSB steganography by exploiting a proper mask inspecting blocks of neighboring bits to detect possible flippings. 

Besides, the work presented by Blasco et al. \cite{httpsan} deals with a proxy for removing hidden data within HTTP headers, but it also contains review of techniques for sanitizing pictures, which are typical in-line objects nested within the hypertext (i.e., the main object). 

Another relevant method is based upon image decompositions algorithms such as wavelets, and then the evaluation of first/higher-order magnitude and phase statistics \cite{hiord}. 

Lastly, we cite the work proposed by Fridrich et al. \cite{fgana}, which deals with compressed images (e.g., {\tt .jpg}) and the F5 algorithm introduced in Section \ref{occ}. Specifically, authors propose to detect the hidden information by estimating the histogram via {\tt 4x4} pixels blocks to compute the number of changes introduced by the F5 algorithm.

\subsubsection{Audio} the detection of LSB-based steganography for audio files is done in a way similar to images (see, e.g., the method proposed in \cite{steansur}). Yet, a more sophisticated approach by Geetha et al. is available in \cite{auste}, which presents a  Genetic Algorithm (GA) classifier to evaluate quality metrics of
audio in order to reveal hidden messages embedded via LSB, frequency-masking and echo hiding. For all cases, the proposed framework proved to be effective, even if hard to implement due to the need of training the classifier with proper datasets. 

Liu et al. \cite{auste2} present a more sophisticated approach. Especially, authors exploit the Mel-Frequency Cepstrum (MFC), i.e., a representation of the short-term power spectrum of a sound able to reveal artifacts. The MFC is then 
 jointly used with the Markov transition features of the second-order derivative to feed a Support Vector Machine (SVM) to recognize the steganographic information. We point out that similar mechanisms are also used for compressed-image, for instance to evaluate secret data injected among the interbands of the DCT domains of {\tt .jpg} carriers, or {\tt .mp3}s. Since the latter are very popular, we mention the work available in  \cite{mp3ste}, where the computation of Markov transition features of each frequency band of an {\tt .mp3} audio is used to recognize hidden data. 

\subsubsection{Video} techniques used for video analysis are similar to those already discussed for images and audio, but also exploiting temporal correlation of the information (i.e., among frames). For the sake of brevity, a more detailed discussion can be found in references \cite{visteg} and \cite{visteg2}, as well as references therein. 

However, an interesting emerging format is the one employed by Youtube, which is one of the most used application  from smartphones and appliances like smart TV or gaming consoles. Thus, in the perspective of performing steganalysis of this format, Zhao et al. \cite{yousteg} use the 3D-DCT to capture the correlation among adjacent frames. Then, statistical features such as absolute central moments, skewness, and kurtosis are used to feed an unsupervised $k$-means clustering classifier. 

\subsubsection{Text} 
as discussed, modern smartphones can produce and manipulate large texts, thus being able to exploit about the totality of linguistic steganography mechanisms available in literature. Similarly, analysis
of related steganograms would require a complete survey, and it is outside the scope of this work (see, reference \cite{stetext} for a thorough discussion on linguistic steganalysis). Recalling that 
smartphones are daily used for editing short text entries (e.g., SMS or emails), or to adjust/view on the fly 
popular formats like {\tt .pdf} documents, we will review the main steganalysis techniques in such fields. Specifically:
%
\begin{itemize}
\item \textit{word spacing}: its detection is done by evaluating the statistical features of spaces in a document. As an example, Li et al. \cite{textan} propose a mechanisms to detect covert data within PDF files;

\item \textit{font formatting}: also in this case, some kind of statistical classifier is trained to detect anomalous font changes or text formatting. For instance \cite{textan2} uses SVM to detect covert channels based on font manipulation, which can be further transmitted via emails or simple {\tt .rtf} files; 

\item \textit{synonyms - translations}: the availability of built-in dictionaries or on-line translators dramatically increases the usage of such techniques. In this perspective, tools to perform steganography-steganalysis of texts have been investigated. This is the case of reference \cite{textan3} dealing with techniques to automatically produce/detect steganograms by means of large  databases of synonyms. 

\end{itemize}
\subsection{Countermeasures Against Local Covert Channels}
The most popular mitigation system engineered for local covert channel was proposed by Enck et al. \cite{td}. Authors introduced \textit{TaintDroid}, i.e., a monitoring system for information flows that operates in real-time in order to detect and prevent leakage of sensitive user data by a misbehaving application. If a flow of information is noticed between a private and public resources, then TaintDroid stops this stream. A key technique utilized is \textit{taint tracking}, which relies on data labeling to enable its tracing and propagation through files, variables and interprocess messages. 

An evolution is QuantDroid \cite{qd}, which is built upon TaintDroid. Besides monitoring contextual information about what data leaves an application, it also quantity the captured ``volume", as a possible metric to reduce the risk of false positives.

Later, an approach called XManDroid (eXtended Monitoring on Android) \cite{xman} has been proposed. It monitors and analyzes communication links across applications by inspecting the inter-component communication calls or the runtime use of the system Application Programming Interface (API). In details, if information flows based on the defined security policy are considered malicious, then they are forbidden. A typical example of a policy is \textit{``an application that has read access to user SMS database must not communicate to an application that has network access}" \cite{xman}.
Therefore, the key design issue of XManDroid (and similar tools) is to define correct security policies, since too tight enforcements  may cause disruptions for non-malicious applications. 

Another similar solution is CHEX \cite{chex}, which instead utilizes static analysis (with the acceptation of off-line). CHEX enables the automatic monitoring of component with vulnerabilities that could be hijacked by a malicious application resulting into sensitive data leakages. To this aim, the tool performs a set of reachability tests on dependence graphs of the smartphone, built by considering the correct/dangerous access control policies among different applications and functionalities. By using this model, the tool is able to evaluate if there is an improperly implemented access granted leading to accidentally leaks of private data, or to execute unauthorized read/write operations on sensitive resources.

A simple solution for mitigating local covert channels for Android smartphones based on modifying volume and vibration settings (e.g., like methods proposed by Schlegel et al. \cite{soundcomber}) has been introduced and evaluated by Hansen et al. \cite{hansen}, and it is advised to be used in conjunction with a system-wide tool like TaintDroid.
Authors propose an application layer exploiting a threshold-based algorithm, considering the past history of events in a time window. In this way, they can detect  burst/anomalous changes in the settings revealing the hidden communication.  By knowing the threshold, malicious applications can beat the countermeasure, but at least they must reduce the transmission rate.

However it must be noted that some of the local covert channels introduced in \cite{collapp}, \cite{wendzel} and \cite{soundcomber} are not detected by TaintDroid (solutions \cite{qd} and  \cite{chex} were not tested since still not publicly available at the time of such studies). As a more
general consideration, we point out that the channels that avoid detection are those reading the ``state" of the OS. Hence, part of the ongoing research is focused on their successful detection.

Lastly, the work of \cite{accelsteg} portraits a countermeasure to defend against covert
channels using the accelerometer sensors available within modern smartphones. In fact,
their increasing resolutions (i.e., $\sim$$100$ MHz) enables the creation of stealth channels
both for transmission and for detecting surrounding events (e.g., the keys entered on a desktop computer if the device is located near - or - on the very same desk). In this perspective, authors propose to implement
a defendant applications randomly producing vibrations to interfere with potential sensing malware.

\subsection{Countermeasures Against Network Covert Channels}

Before reviewing countermeasures for network covert channels, we point out that,
as today, there are not any solutions solely developed for smartphones. Yet, they were proposed for generic TCP/IP architectures, hence they can be also applicable to mitigate network steganography in smartphones (even if with some minor tweaks).

Generally, potential mitigation techniques against network covert channels can be broadly classified into one of \textit{three} groups:
\begin{itemize}
\item \textit{network traffic normalizers}: normalization aims at removing ambiguities in network traffic that is exploited for steganographic schemes (e.g., the alteration of inter packet times to encode secret data). 
Typically, normalizers are divided according to whether they capture a context of the transmission, thus  \cite{lucena}: 
\begin{itemize}
\item \textit{stateless}: they analyze only one packet at time;
\item \textit{statefull}: they consider a packet in a wider context, e.g., 
an end-to-end conversation. 
\end{itemize}

Existing normalizers can be a part of the intrusion detection/prevention systems like Snort \cite{snort} or a dedicated solution like the network-aware active warden proposed in \cite{netwarden} or the network pump \cite{pump}. 
In a similar extent, Lewandowski et al. \cite{ipv6st} deal with general techniques applied to IPv6, which will be increasingly adopted in future 5G and beyond mobile devices.

The biggest drawbacks of network normalizers include \cite{tn}: \textit{i)} side-effects as due to normalization process, e.g., part of the protocols headers cannot be further utilized, thus causing a loss of functionality; \textit{ii)} the limited buffer size used to perform analysis could be exhausted when in presence of large traffic volumes, hence they lack of scalability; \textit{iii)}  packets can be sent using different routes resulting in incomplete normalizations. As a workaround, traffic should be monitored in different portion of the network, eventually resulting in a framework difficult to configure and maintain;  \textit{iv)}  traffic normalization cannot be too aggressive as to avoid degradation to Quality of Service (QoS) or Quality of Experience (QoE), hence it is not a definitive solution against covert channels; 

\item \textit{statistical methods} include approaches to reveal/prevent encoding secret data bits into features characterizing the network traffic (e.g., the inter-arrival time of PDUs). As paradigmatic cases, Berk et al. \cite{berk} focus on techniques comparing the inter-arrival times distributions of ``clean" network traffic and steganographically modified one, while in \cite{gian} the entropies of packets' inter-arrival times are inspected and then matched. 

\textit{Three} another approaches were proposed by Cabuk et al. \cite{cabuk}. The main one exploits data compression applied to inter-arrival time values as the indicator of the presence of a covert channel. Hence: \textit{i)} inter-arrivals are computed for each packet; \textit{ii)} values are 
converted as strings; \textit{iii)} the resulting ``text" is compressed. The more 
the compression is efficient, the greater is the risk that the flow of packets contains a covert channel. In fact, the alteration of inter-arrival times to hide data would produce more regular timing statistics, thus making the compression algorithm to perform better (i.e., equal times will lead to equal strings). The remaining two statistical methods are based on: comparing standard deviation of recorded inter-arrival times, and evaluation of $\epsilon$-similarity; 

\item \textit{machine learning approaches} rely mainly on trained classifiers  capable of distinguishing between traffic with a covert channel and normal network flows. Such solutions include utilization of SVMs \cite{sohn}, neural networks \cite{tumoian} or C$4.5$ decision trees \cite{zander2} and \cite{zander3}. Similarly for the case of object covert channels, also such methods have the main drawback of needing a proper dataset for training the classifiers. 
\end{itemize}

\section{Conclusions and Future Research Directions}
\label{cfw}
Modern smartphones are an excellent playground for the development of new steganographic methods,
and also, as a consequence of their importance, they will be likely to become one of the most targeted
platforms for data exfiltration. To classify the work done, we developed a taxonomy to partition methods in three covert channels, and we also evaluated possible countermeasures. We can summarize the lessons learned and the major findings of this survey as follows: 
\begin{itemize}
\item currently the most active trend is the one aiming at the creation of local covert channels for  Android-based devices. By considering the future developments of many works, it appears that in the near future new low-bandwidth methods will be introduced, as well as the willingness of the research community of starting
some active development over iOS or WindowsPhone;

\item currently little attention is devoted to network covert channels exploiting specificities of  smartphones. Yet,  this field is the most promising research one for smartphone steganography, especially because routing data outside the device secretly is a core requirement to make the colluding applications scenario a real threat;

\item digital media steganography for smartphones does not add any novelty compared to desktops, apart some works on QR codes. Nevertheless, video steganography applied to mobile devices is still a largely underestimated research area;

\item the tools ``for the masses" available in application markets do not appear as enablers for serious, hardly discoverable secret communications. Yet, they allow to present the concept of information hiding to the laymen;

\item a portion of the mitigation techniques addresses some of the smartphone steganography methods (mostly those that result in local covert channel), but they are not universal enough  nor efficient to be practically deployed for real-time detection;

\item as it can be noticed by the differences in the amount of papers surveyed, the research community is more active in developing covert channels, rather than countermeasures. In addition, 
the latter are method-specific and there is not any general solution especially for multidimensional carriers as smartphones.

\end{itemize}

We hope that this survey motivates to increase the efforts on researching both steganography and steganalysis over smartphones. By evaluating the pros and cons of each paper, our investigation revealed some uncovered areas. Therefore, in our perspective, some potential research directions are: 
\begin{itemize}
\item \textit{energy consumption}: investigating the energy consumptions of networking and security mechanisms is a hot research topic, especially on handheld devices \cite{miogreen}. Alas, about the totality of surveyed  works neglect this particular. Moreover, understanding the impact of steganography in terms of power depletion can be used as an effective detection method. Part of the future research should be aimed at filling this gap;

\item \textit{learn from the past}: even if they could appear anachronistic, SMS-like steganography could be a relevant source of data and hazards in the near future. In fact, the availability of flat-rate 3G data plans, jointly with applications for sending messages through the Internet (e.g., Apple iMessage, Viber or Whatsapp) will likely to be affected by the same mechanisms,  but on a magnified scale factor. This could be a challenging scenario especially for detection, which should be done in real-time and scalable manner, also by preserving privacy;

\item \textit{devices and clouds}: the diffusion of the cloud-based paradigm enables a variety of services moving data from the device to the Internet (and vice versa). In this vein, a plethora of new patterns should be investigated as possible steganography carriers, e.g., crafted subsequent syncing of documents (for instance when collaborating through shared storage like Google Drive), data embedment within voice-based services (for instance, Google Voice) and protocol hijacking of Web-based GUIs. At the same time, this field of research also needs the development of proper countermeasures;

\item \textit{develop mitigation techniques}: as shown, local covert channels (Section \ref{lcc})  are still
poorly detected, especially those using the ``state" of the OS. Thus, this field of research would be a very challenging playground in the near future. 
\end{itemize}

\section*{ACKNOWLEDGMENTS}
This research was partially supported by: the Polish National Science Center under Grant N. 2011/01/D/ST7/05054, and by the European Union in the framework of European Social Fund through the Warsaw University of Technology Development Programme.

We thank the anonymous Reviewers, whose comments allowed to increase the quality of this work.

\vspace{2cm}

\vspace{10cm}
\begin{IEEEbiography}[{\includegraphics[width=1in,
height=1.25in,clip, keepaspectratio]{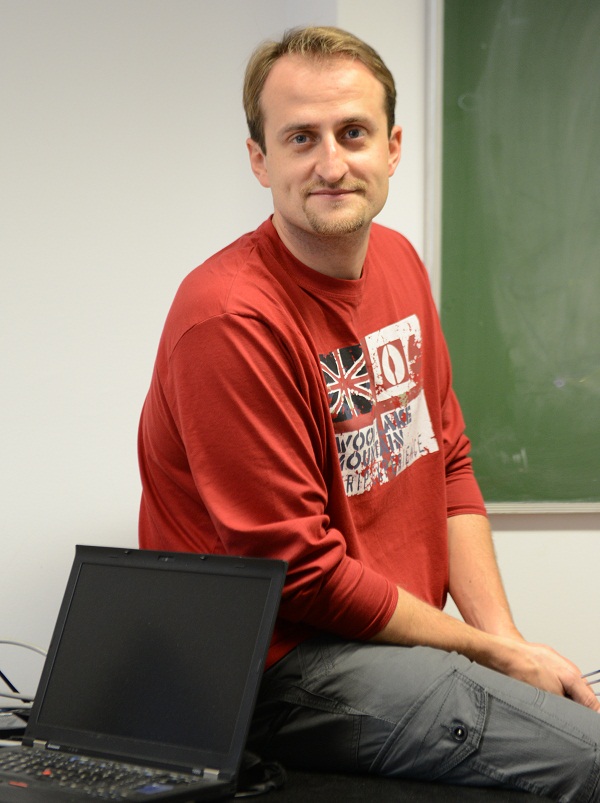}}]{Wojciech Mazurczyk}
holds B.Sc. (2003), M.Sc. (2004), Ph.D. (2009, with honours) and D.Sc. (habilitation, 2014) all in Telecommunications from Warsaw University of Technology (WUT), Poland, where currently he is associate professor. He is the author of more than 80 scientific papers, 1 patent application, and more than 30 invited talks on information security and telecommunications. He is a Research co-leader of Network Security Group ({\tt secgroup.pl}), and his main research interests are: 
information hiding techniques, network anomalies detection, digital forensics, network security and multimedia services. Besides, he is a TPC member of a number of refereed conferences, including IEEE INFOCOM, IEEE GLOBECOM, IEEE ICC and ACSAC. He also serves as the reviewer for a number of major refereed international magazines and journals. From 2013 he is an Associate Technical Editor for the IEEE Communications Magazine, IEEE Comsoc. Personal website: {\tt http://mazurczyk.com}. 
\end{IEEEbiography}

\vspace{-12cm}

\begin{IEEEbiography}[{\includegraphics[width=1in,
height=1.25in,clip, keepaspectratio]{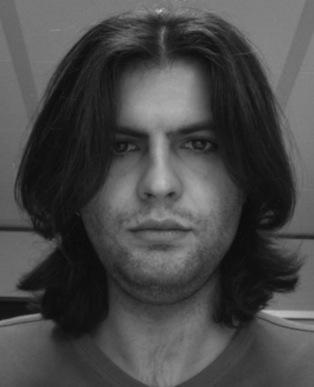}}]{Luca Caviglione}
is a Researcher at the Istituto di Studi sui Sistemi Intelligenti per l'Automazione (ISSIA) of the National Research Council of Italy (CNR). In 2007 he was with the Italian National Consortium for Telecommunications (CNIT), University of Genoa Research Unit. He has a PhD in Electronic and Computer Engineering from the University of Genoa, Italy. His research interests include peer-to-peer (p2p) systems, IPv6, social networks, wireless communications, and network security. He is author and co-author of more than 90 academic publications, and several patents in the field of p2p. He has been involved in Research Projects funded by the ESA, the EU and MIUR. He is a Work Group Leader of the Italian IPv6 Task Force, a contract Professor in the field of p2p networking and a Professional Engineer. He is involved in the Technical Program Committee of many International Conferences, and regularly serves as a reviewer for the major International Journals. From 2011, he is an Associate Editor for the Transactions on Emerging Telecommunications Technologies, Wiley. 
\end{IEEEbiography}

\end{document}